\documentclass[aps,prd,onecolumn,superscriptaddress,nofootinbib]{revtex4}
\usepackage{epsfig,amsfonts,amsthm}
\usepackage{amsmath,amssymb,bbold}
\usepackage{color}
\usepackage[utf8]{inputenc}
\renewcommand{\Re}{\mathrm{Re }}
\renewcommand{\Im}{\mathrm{Im }}
\newcommand{\doublet}[2]{ \left( \begin{array}{c}#1 \\ #2 \end{array}\right) }

\newcommand{\lr}[1]{ \langle #1 \rangle}

\newcommand{\Z}{\mathbb{Z}}

\providecommand{\mtrx}[1]{\begin{pmatrix} #1 \end{pmatrix}}

\newcommand{\id}{\mathbb{1}}
\newcommand{\lrpartial}{\,\partial^{\hspace{-7pt}\raise3pt\hbox{\small $\leftrightarrow$}}\!}
\newcommand{\scr}[1]{\mbox{\scriptsize #1}}

\def\lsim{\mathrel{\rlap{\lower4pt\hbox{\hskip1pt$\sim$}}
    \raise1pt\hbox{`$<$}}}         %less than or approx. symbol
\def\gsim{\mathrel{\rlap{\lower4pt\hbox{\hskip1pt$\sim$}}
    \raise1pt\hbox{$>$}}}         %greater than or approx. symbol

\begin{document}
%\title{Phase diagram and phase transitions in the $\Sigma(36)$ 3HDM}
\title{Charge-breaking domain walls separating neutral vacua in multi-Higgs models}

\author{Yucheng Yang}\thanks{E-mail: yangych63@mail2.sysu.edu.cn}
\affiliation{School of Physics and Astronomy, Sun Yat-sen University,
519082 Zhuhai, China}
\author{Igor P. Ivanov}\thanks{E-mail: ivanov@mail.sysu.edu.cn}
\affiliation{School of Physics and Astronomy, Sun Yat-sen University,
519082 Zhuhai, China}

\begin{abstract}
The scalar potential of a multi-Higgs model can possess 
a rich structure of minima and saddle points,
which evolves in an intricate way as the parameters change.
In the hot early Universe, it could trigger multi-step phase transitions, 
with exotic intermediate phases and peculiar domain wall configurations.
Here, we provide a glimpse into this richness with the example of the three-Higgs-doublet model
with the symmetry group $\Sigma(36)$, either exact or softly broken.
We present its phase diagram tracking not only the global minimum but all of its extrema.
In particular, we reveal parameter space regions in which the deepest saddle point
is charge breaking. This naturally leads to 
phase transitions between neutral vacua
which involve expanding and colliding charge-breaking bubble walls.
We also comment on opportunities of multi-step phase transitions,
on charge-breaking intermediate phases, 
and on phase transitions between different charge-breaking vacua.
We illustrate the general discussion with two benchmark models,
one of which possesses competing saddle points and allows 
for emergence of bubbles of the same true vacuum but with bubble walls of different nature.
The intriguing cosmological implications of all these possibilities deserve dedicated study.

\end{abstract}

\maketitle

\section{Introduction}

\subsection{Phase transitions in multi-Higgs models}

Non-minimal Higgs sectors are popular when building models beyond the Standard Model (SM),
see the recent reviews \cite{Branco:2011iw,Kanemura:2014bqa,Ivanov:2017dad,Arcadi:2019lka}.
With very few assumptions, such models can address shortcomings of the SM 
and offer a surprisingly rich list of collider, neutrino physics, astroparticle, and cosmological signatures.
A particular aspect of multi-Higgs models is a rich variety of phase transitions 
in early Universe they can induce, see \cite{Hindmarsh:2020hop,Athron:2023xlk} for recent reviews.
At finite temperature $T$, the multi-Higgs potential can possess several competing minima,
whose positions and depths evolve with $T$ in a non-trivial way. 
Depending on the number of Higgs fields and the details of their interaction,
this evolution can lead to multi-step phase transitions, which can be 
observed even within the simplistic tree-level high-$T$ approximation of the two-Higgs-doublet model (2HDM) \cite{Ginzburg:2009dp,Ivanov:2008er,Ginzburg:2010wa}.
In the past few years, multi-step phase transitions were studied in many papers 
within the loop-corrected effective potential, with a focus on baryogenesis and 
generation of non-standard spectra of gravitational waves,
see for example \cite{Vaskonen:2016yiu,Chala:2018opy,Morais:2019fnm,Aoki:2021oez,Benincasa:2022elt,Cao:2022ocg,Aoki:2023xnn,Aoki:2023lbz,Basler:2024aaf}
and references therein.

Phase transition sequences can include unusual intermediate phases
and lead to other surprises.
One example is electroweak symmetry non-restoration at high temperatures,
a phenomenon known since long ago \cite{Weinberg:1974hy} 
and recently demonstrated within the 2HDM \cite{Aoki:2023lbz,Basler:2024aaf}.
Another important feature is the possibility of trapped vacua,
that is, the situation when the Universe stays in a local metastable minimum 
all the way down to $T\to 0$ just because the tunneling probability never gets sufficiently 
high to trigger bubble nucleation \cite{Biekotter:2021ysx,Biekotter:2022kgf}.

Yet another intriguing possibility is the role of charge-breaking (CB) Higgs field configurations \cite{Barroso:2006pa,Ginzburg:2007jn}.
As pointed out in \cite{Ginzburg:2009dp,Ivanov:2008er}, in a certain parameter space region of the 2HDM, 
the Universe, while cooling down, could have evolved through an intermediate CB phase.
This possibility was recently put under scrutiny in \cite{Aoki:2023lbz}.
Treating the finite-$T$ one-loop effective potential of the 2HDM with the aid of the code {\tt BSMPTv2}
\cite{Basler:2020nrq},
the authors found many benchmark points which exhibited charge-breaking phase at intermediate temperatures
and which, at the same time, satisfied the present day collider constraints on the Higgs sector.
Moreover, temperature evolution of certain benchmark points of \cite{Aoki:2023lbz} 
exhibited CB-to-CB first-order phase transitions within the charge-breaking phase.
The very recent update of the code, {\tt BSMPTv3}, confirmed the existence of intermediate CB phases 
in the evolution of the benchmark 2HDMs \cite{Basler:2024aaf}.

Another remarkable feature of multi-Higgs potentials 
is the possibility of domain walls separating space-time domains of different minima, as well as other topologically stable configurations,
which can arise from spontaneous breaking of global symmetries \cite{Zeldovich:1974uw,Kibble:1976sj}.
If completely stable, these extended objects face severe cosmological constraints.
But if they arise only temporarily, in the course of multi-step phase transitions, domain walls can play crucial role 
in electroweak phase transition by seeding and exponentially enhancing nucleation of the true vacuum bubbles,
see for example the recent studies \cite{Blasi:2022woz,Agrawal:2023cgp} and references therein.

A new twist on this topic appears when two domains of neutral minima are separated by a domain wall
which is charge-breaking. The structure of domain walls linking different minima 
was studied through detailed numerical modeling in \cite{Battye:2020sxy,Battye:2020jeu}.
It was indeed found that the path in the Higgs space providing smooth field interpolation between the two minima
could contain a region with slight departure form the neutrality condition.
However, the CB effect observed in \cite{Battye:2020sxy,Battye:2020jeu} was minor.

However it is well possible to construct a 2HDM Higgs potential featuring two neutral minima separated by the deepest
saddle point which is charge breaking.
This is a rather straightforward exercise, and it is surprising that, for a long time, it was not discussed
in literature. Only when this paper was in preparation, the publication \cite{Sassi:2023cqp} appeared,
in which such CB domain walls in the 2HDM were investigated in great detail, 
including fermion scattering off a CB wall.
%We stress that, when the interpolating Higgs field configuration goes through a low-lying CB saddle point,
%departure from the neutrality condition is very sizable.

If the potential has two minima at different depths,
the Universe residing initially in the false vacuum state can tunnel into the true vacuum.
This transition proceeds by nucleation of bubbles of true vacuum, which then expand
in the false vacuum background.
The charge breaking bubble walls sweep through the primordial hot plasma and eventually collide, 
leading to bubble merging until the true vacuum region covers the entire Universe.
Although the extensive numerical simulations of \cite{Sassi:2023cqp} offer some insights into this process,
we believe that this novel regime of thermal evolution deserves a closer study,
as it may offer new opportunities for baryogenesis and even generation of primordial magnetic fields
\cite{Durrer:2013pga,Subramanian:2015lua,Olea-Romacho:2023rhh}.

Although the domain walls and intermediate phases, including charge-breaking configurations, exist within the 2HDM,
we would like to check what novel features appear in even richer scalar sectors.
For example, the three-Higgs-doublet model (3HDM), which was originally introduced by S.~Weinberg in 1976 \cite{Weinberg:1976hu}, 
is now being actively studied. The interest in the 3HDM is mostly driven by its richness unseen in the 2HDM,
both in terms of symmetry content and model-building opportunities in the scalar and Yukawa sectors,
and in terms of various phenomenological and astroparticle signals which were not compatible in the 2HDM but 
which coexist with three Higgs doublets, see \cite{Ivanov:2017dad} for a review.
In particular, it is known that the 3HDM potential can support several minima, 
including coexisting neutral and CB minima \cite{Ivanov:2010wz}, which may
evolve with temperature in different ways. 
This evolution may lead, even at the tree level, to longer multi-step sequences of phase transitions 
with unusual intermediate phases or important roles played by domain walls, to novel opportunities for trapped vacua, 
and to overlapping or even competing phase transitions.
All these phenomena and their possible observable consequences such as gravitational wave production
and baryogenesis are worth exploring.
It goes without saying that no systematic study exists of these issues within the 3HDM.

\subsection{Goals of the present work}

In this paper, we would like to stimulate this research program by considering 
the phase diagram and phase transitions possible within a specific version of the 3HDM
based on the symmetry group $\Sigma(36)$. 
This choice is motivated by the fact that 
$\Sigma(36)$ is the largest finite symmetry group
which can be imposed on the scalar sector of the 3HDM without leading to accidental continuous symmetries
\cite{Ivanov:2012fp,Ivanov:2012ry}. Its scalar sector contains only four free parameters, 
only two of them being structurally important.
This will allow us to visualize the phase diagram and to study many of its aspects analytically, 
including some consequences of soft breaking terms.
Even in this simple setting, we can observe such phenomena as charge-breaking domain walls
linking neutral minima, and CB-to-neutral and CB-to-CB phase transitions as we vary the coefficients.

We stress that the $\Sigma(36)$ 3HDM is used here only as a toy model. We limit ourselves to the scalar sector and do not try to fit experimental data nor to include fermions. The main objectives of this paper are to explore the structural richness of multi-Higgs potentials beyond just looking at the global minima and to demonstrate that even in this constrained setting one find structural phenomena that have not received much attention. We believe that these features and the patterns of phase transitions are typical for multi-scalar potentials and may arise in phenomenologically viable multi-Higgs models beyond the 2HDM. We hope that the lessons learned here will guide model-builders in constructing multi-Higgs models with exotic but custom-tailored cosmological history.

The structure of this paper is as follows.
In Section~\ref{section:general-remarks} we review the general properties of the $\Sigma(36)$ 3HDM, 
describe its phase diagram, and discuss its possible global minima.
Section~\ref{section:numerical} presents the surprisingly rich structure of extrema of the potential as a function of free parameters.
We first show the results for the exact $\Sigma(36)$ symmetry group. Then we add soft breaking terms
which preserve some of the symmetries and explore how the structure of minima and extrema evolves 
as the soft breaking parameters change. 
In Section~\ref{section:domain}, we explore in more detail the specific issue of charge-breaking domain walls, 
first in the 2HDM and then in our toy model.
We also give here two benchmark models which highlight two different regimes of the thermal evolution
in the softly broken $\Sigma(36)$.
Finally, in Section~\ref{section:discussion} we draw our conclusions and suggest questions for follow-up studies.
Additional technical results of our numerical study are presented in the Appendix.

\section{$\Sigma(36)$ 3HDM: general properties}\label{section:general-remarks}

\subsection{The scalar sector of the multi-Higgs-doublet models}\label{subsection:NHDM}

In the $N$-Higgs-doublet model, we assume that the gauge interactions and the fermion content is
exactly the same as in the SM and extend only the scalar sector by adopting the idea of several Higgs generations.
We introduce $N$ Higgs doublets $\phi_i$, all of them having the same gauge quantum numbers,
which can couple to the fermions via the standard Yukawa interactions and, importantly,
interact with themselves.
Their self-interaction is described by renormalizable scalar self-interaction potential 
which can be generically written as 
\begin{equation}
	V = m_{ij}^2 \phi_i^\dagger \phi_j + \lambda_{ijkl}(\phi_i^\dagger \phi_j)(\phi_k^\dagger \phi_l)\,,\label{V-NHDM}
\end{equation}
all indices going from 1 to $N$. In the general NHDM, the parameters $m_{ij}^2$ and $\lambda_{ijkl}$ 
are only constrained by the requirement that the potential is hermitian,
which leaves room for a very large number of free parameters and renders the model hardly predictable.
However, if the model is equipped with additional global symmetries forming the symmetry group $G$, 
then only a few parameters are allowed, and the model can acquire characteristic phenomenological features
\cite{Ivanov:2017dad}.

Let us now focus on the 3HDM.
Once the Higgs potential is written, one must find its global minimum and determine 
the vacuum expectation value (vev) alignment of all scalar doublets $\lr{\phi_i}$:
\begin{equation}
	\lr{\phi_1} = \doublet{u_1}{\frac{1}{\sqrt{2}}v_1}\,,\quad  
	\lr{\phi_2} = \doublet{u_2}{\frac{1}{\sqrt{2}}v_2}\,,\quad 
	\lr{\phi_3} = \doublet{u_3}{\frac{1}{\sqrt{2}}v_3}\,,
	\label{general-vevs}
\end{equation}
where all vev components can, in principle, be complex. 
Usually, one employs the global $SU(2)\times U(1)$ transformation to set $u_1 = 0$
and make $v_1$ real, but in this paper, we will not always adhere to this procedure. 
In the perturbative treatment of the model, the vev alignment defines the vacuum state;
by expanding the lagrangian around the vacuum and diagonalizing the mass matrices, 
one obtains the physical states and their interactions.

There are two general classes of vacuum points: neutral vacuum and charge breaking vacuum.
A neutral vacuum corresponds to the normal situation: the electroweak gauge group $SU(2)_L \times U(1)_Y$
breaks down to the electromagnetic gauge group $U(1)_{em}$, the photon remains massless,
anmd the electric charge is conserved.
For a neutral vacuum, there is always a global $SU(2)\times U(1)$ transformation 
which sets all the upper components $u_i = 0$, so that the vev alignment becomes
\begin{equation}
	\lr{\phi_1} = \frac{1}{\sqrt{2}}\doublet{0}{v_1}\,,\quad  
	\lr{\phi_2} = \frac{1}{\sqrt{2}}\doublet{0}{v_2}\,,\quad 
	\lr{\phi_3} = \frac{1}{\sqrt{2}}\doublet{0}{v_3}\,,
	\label{neutral-vevs}
\end{equation}
with $v_i$ being real or complex.
If this is the case, we will often take the overall scale out, implicitly assuming that 
$|v_1|^2 + |v_2|^2 + |v_3|^2 = v^2$, where $v = 246$~GeV, 
and denote the vev alignment
just by indicating the ratios between individual vevs. 
For example, one of the minima shown below (point $C_1$) has the vev alignment $v_1 = v_2 = v_3$.
Denoting it as $v_C$, we write
\begin{equation}
	(\lr{\phi_1^0},\, \lr{\phi_2^0},\, \lr{\phi_3^0}) = 
	\left(\frac{v_C}{\sqrt{2}},\, \frac{v_C}{\sqrt{2}},\, \frac{v_C}{\sqrt{2}}\right) = \frac{v_C}{\sqrt{2}}\left(1, 1, 1\right)\,,
	\quad
	\mbox{with}\quad v_C = \frac{v}{\sqrt{3}}\,,
\end{equation}
and label this alignment as $(1,1,1)$.

If the vacuum is charge-breaking, then some of the upper components $u_i$ remain non-zero
in all bases. In this situation, the electroweak gauge group is broken completely, 
and all gauge bosons acquire masses. Although this situation does not correspond to vacuum we observe now,
it could be a viable option for the early hot Universe in a finite range of temperatures, 
see the recent 2HDM study in \cite{Aoki:2023lbz,Basler:2024aaf}.

One important remark concerns the choice of vev alignment. In the analysis of multi-Higgs models,
it has become a standard practice to select a desired vev alignment as input parameters 
and express some of the parameters of the scalar potentials in terms if vevs.
We draw the reader's attention to the fact that in multi-Higgs models with large discrete 
symmetry groups it is not always possible. Namely, minimization condition of such a scalar potential
can only be satisfied at very special vev alignments with relations among $v_i$,
which are stable against continuous variation of the parameters of the potential.
A particular consequence of this phenomenon is that a large discrete symmetry group $G$
of the starting potential is not broken completely. 
In any of the possible minima, some of the original symmetries are still preserved,
while the other, broken symmetries, link several distinct minima of the potential.
In the case of the 3HDMs equipped with finite symmetry groups, all the patterns of symmetry breaking
for all possible minima were classified in \cite{Ivanov:2014doa}.

\subsection{The potential and its parameters}\label{subsection:potential}

The symmetry group $\Sigma(36)$ was first identified as a viable option for the 3HDM in \cite{Ivanov:2012ry,Ivanov:2012fp}.
Group-theoretically, it is defined as a $\Z_4$ permutation acting on the generators of the abelian group 
$\Z_3\times \Z_3$: $\Sigma(36) \simeq (\Z_3 \times \Z_3)\rtimes \Z_4$.
The elements of this group act on the three doublets by unitary matrices: $\phi_i \mapsto g_{ij} \phi_j$,
where each $g$ can be expressed as products of powers of the generators of the group.
In a suitable basis, these generators, denoted as $a$, $b$, and $d$, have the following expressions:
\begin{equation}
	a = \mtrx{1&0&0\\ 0&\omega&0\\ 0&0&\omega^2}\,, \quad
	b = \mtrx{0&1&0\\ 0&0&1\\ 1&0&0}\,,\quad
	d = \frac{i}{\sqrt{3}} \left(\begin{array}{ccc} 1 & 1 & 1 \\ 1 & \omega^2 & \omega \\ 1 & \omega & \omega^2 \end{array}\right)\,,
	\label{Sigma36-generators}
\end{equation}
where $\omega = \exp(2\pi i/3)$. The orders of these generators are:
$a^3 = \id$, $b^3 = \id$, $d^4 = \id$.
Note that $d^2$ interchanges $\phi_2$ and $\phi_3$; 
thus, $\Sigma(36)$ contains all permutations of the three doublets.
Had we imposed symmetry under $d^2$ but not $d$, 
we would end up with the more familiar symmetry group $\Delta(54)$,
first used within the 3HDMs in 1970s \cite{Segre:1978ji} 
and explored later in \cite{Grimus:2010ak, Merle:2011vy,Hagedorn:2013nra, Rong:2016cpk}.
It is interesting that the $\Sigma(36)$ 3HDM has not yet received 
a detailed phenomenological study; only its main structural features are known
\cite{Ivanov:2014doa,deMedeirosVarzielas:2021zqs}.
For more details on the relation between $\Delta(54)$ and $\Sigma(36)$ and the subtleties of their definitions, 
see \cite{Ivanov:2012fp}; a concise version of this discussion can be found in Appendix~A of Ref.~\cite{deMedeirosVarzielas:2021zqs}.

The 3HDM scalar potential invariant under $\Sigma(36)$ has the following form:
\begin{eqnarray}
	V_0 & = &  - m^2 \left(\phi_1^\dagger \phi_1+ \phi_2^\dagger \phi_2+\phi_3^\dagger \phi_3\right)
	+ \lambda_1 \left(\phi_1^\dagger \phi_1+ \phi_2^\dagger \phi_2+\phi_3^\dagger \phi_3\right)^2 \nonumber\\
	&&
	- \lambda_2 \left[|\phi_1^\dagger \phi_2|^2 + |\phi_2^\dagger \phi_3|^2 + |\phi_3^\dagger \phi_1|^2
	- (\phi_1^\dagger \phi_1)(\phi_2^\dagger \phi_2) - (\phi_2^\dagger \phi_2)(\phi_3^\dagger \phi_3)
	- (\phi_3^\dagger \phi_3)(\phi_1^\dagger \phi_1)\right] \nonumber\\
	&&
	+ \lambda_3 \left(
	|\phi_1^\dagger \phi_2 - \phi_2^\dagger \phi_3|^2 +
	|\phi_2^\dagger \phi_3 - \phi_3^\dagger \phi_1|^2 +
	|\phi_3^\dagger \phi_1 - \phi_1^\dagger \phi_2	|^2\right)\, .
	\label{Vexact}
\end{eqnarray}
The first two lines here are invariant under all $SU(3)$
transformations of the three Higgs doublets,
%The positive sign of $\lambda_2$ guarantees that the minimum corresponds to a neutral vacuum,
%but the minimization of these two lines alone would lead to several neutral Nambu-Goldstone bosons.
and it is only the $\lambda_3$ term that selects the finite group $\Sigma(36)$ out of the entire $SU(3)$.
The potential in Eq.~\eqref{Vexact} is also $CP$ invariant. In fact, as was proved in \cite{Ivanov:2011ae,Ivanov:2014doa},
the presence of the $\Z_4$ symmetry group within the 3HDM scalar sector automatically forbids 
any form of $CP$ violation in the scalar sector, be it explicit or spontaneous.

All four free parameters in Eq.~\eqref{Vexact} are real but play different roles.
The coefficients $m^2$ and $\lambda_1$ only fix the overall scales 
of the vacuum expectation value $v$ and the SM-like Higgs mass,
but play no role in shaping the structure of the potential and in defining the number and locations of its extrema. 
This leaves us with only two structurally important coefficients: $\lambda_2$ and $\lambda_3$
(to be more accurate, $\lambda_2/\lambda_1$ and $\lambda_3/\lambda_1$).
This is why when we study the structural features of the model we will fix $m^2$ and $\lambda_1$
and explore different regions of $\lambda_2$ and $\lambda_3$.
Let us also mention right away that, in order for the potential to be bounded from below (BFB), 
the quartic parameters must satisfy the following constraints:
\begin{equation}
	\lambda_1 > 0\,, \quad \lambda_1 + \lambda_3 > 0\,,\quad
	\lambda_1 + \frac{1}{4}\lambda_2 > 0\,,\quad \lambda_1 + \frac{1}{4}(\lambda_2 + \lambda_3) > 0\,.
	\label{BFB}
\end{equation}
Note that all these inequalities are strict, which means that we impose the BFB conditions in the strong sense \cite{Maniatis:2006fs}.
It is known that, in a generic multi-Higgs model, one can often resort to the weaker BFB conditions on the quartic couplings 
which include equalities. In such a situation, the quartic part of the potential contains flat directions, 
but the potential is still bounded from below if the quadratic terms frow along such directions.
However, in our case, the quadratic potential in Eq.~\eqref{Vexact} is always negative, 
which forces us to insist on the strong BFB conditions.

\subsection{The global minima}

An important feature of the entire $\Delta(54)$ family of the 3HDM models, including the $\Sigma(36)$ 3HDM, is the ``rigidity'' of its global minima upon continuous variations of the parameters, a feature which is known since 1984 \cite{Branco:1983tn}. The best way to see it is to employ the geometric method developed in \cite{Degee:2012sk,Ivanov:2014doa}.
We first define
\begin{eqnarray}
	r_0 &=& \frac{1}{\sqrt{3}}\left(\phi_1^\dagger \phi_1+ \phi_2^\dagger \phi_2+\phi_3^\dagger \phi_3\right)\,,\nonumber\\
	x &=& 1 - \frac{z_{12} + z_{23} + z_{31}}{r_0^2}\,,\quad \mbox{where}\quad
	z_{ij} = (\phi_i^\dagger \phi_i)(\phi_j^\dagger \phi_j) - |\phi_i^\dagger \phi_j|^2\nonumber\\
	y &=& \frac{|\phi_1^\dagger \phi_2 - \phi_2^\dagger \phi_3|^2 + |\phi_2^\dagger \phi_3 - \phi_3^\dagger \phi_1|^2 + 
		|\phi_3^\dagger \phi_1 - \phi_1^\dagger \phi_2|^2}{3 r_0^2}\,,\label{r0xy}
\end{eqnarray}
and then rewrite the scalar potential as 
\begin{equation}
	V_0 = - \sqrt{3}\, m^2 r_0 + r_0^2\left(\Lambda_1 + \Lambda_2 x + \Lambda_3 y\right)\,,\label{linear-form}
\end{equation}
where $\Lambda_1 = 3\lambda_1 + \lambda_2$, $\Lambda_2 = - \lambda_2$, $\Lambda_3 = 3 \lambda_3$.
In this way, we separate the ``radial'' and ``angular'' steps of the minimization problem.
We first minimize $\Lambda(x,y) \equiv \Lambda_1 + \Lambda_2 x + \Lambda_3 y$ as a function of $x$ and $y$, 
and then, having found its minimal value $\tilde\Lambda$, we minimize the potential with respect to $r_0$,
which yields
\begin{equation}
	r_0 = \frac{\sqrt{3}\,m^2}{2 \tilde\Lambda}\,, \quad V_{\scr{min}} = - \frac{3 m^4}{4 \tilde\Lambda}\,.
	\label{general-min}
\end{equation}
The $SU(3)$ invariant quadratic term automatically leads to scalar alignment,
so that we obtain a SM-like Higgs boson with the mass $m_h^2 = 2m^2$.
The masses of the other scalars exhibit remarkable degeneracy patterns \cite{deMedeirosVarzielas:2022fnn}.

\begin{figure}[!htb]
	\centering
	\includegraphics[height=5cm]{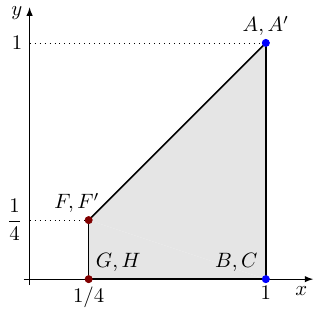}
	\qquad\qquad
	\includegraphics[height=5cm]{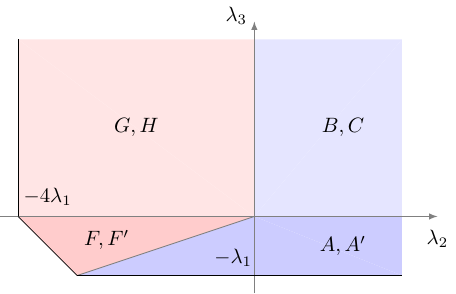}
	\caption{Global minima of the $\Sigma(36)$-symmetric 3HDM. 
		Left: the orbit space in the $(x,y)$ plane defined by Eqs.~\eqref{r0xy}. Right: the $(\lambda_2,\lambda_3)$ plane 
	bounded by the BFB conditions in Eqs.~\eqref{BFB}, with the global minima indicated in each region 
	(neutral and charge-breaking minima shown in blue and in pink, respectively).}
	\label{fig-orbit-space}
\end{figure}

What makes the form of the scalar potential in Eq.~\eqref{linear-form} unique is the linear dependence of $V_0$ on $x$ and $y$. 
In order to find $\tilde\Lambda$, we just need to describe the orbit space, 
that is, the shape in the $(x,y)$ plane formed by all possible Higgs doublet configurations,
and identify its vertices.
In Fig.~\ref{fig-orbit-space}, left, we show this orbit space. It has the trapezoid shape bounded by the 
inequalities $1/4 \le x \le 1$ and $0 \le y \le x$, for the derivation see Ref.~\cite{Ivanov:2014doa}.
The value of $x$ becomes especially clear in the bilinear formalism, which we briefly review in Appendix~\ref{appendix:bilinear} and which we will use at several instances in this paper.
The boundary segment corresponding to $x = 1$ represents the neutral vacua \cite{Ivanov:2010ww}.
This can be easily seen from Eq.~\eqref{r0xy}, as a neutral vacuum point implies that all $z_{ij} = 0$.
The bulk of the orbit space corresponds to the charge breaking vev configurations.

Clearly, $\tilde\Lambda$ must be positive everywhere in the orbit space. 
Writing down the condition $\tilde\Lambda > 0$ 
at the four vertices of the trapezoid leads to the BFB conditions listed in Eq.~\eqref{BFB}. 

Minimizing a linear function of $x$ and $y$ in a polygon is elementary.
Depending on the parameters, $\tilde\Lambda$ achieves its minimum
at one of the four vertices\footnote{If the point $(\lambda_2,\lambda_3)$ lies exactly 
on a boundary between two regions, then the global minimum is attained along an entire segment 
linking the two corresponding vertices in the orbit space.}, 
labeled in Fig.~\ref{fig-orbit-space}, left.
The corresponding regions on the $(\lambda_2,\lambda_3)$ plane are shown in Fig.~\ref{fig-orbit-space}, right.
Below we list these four situations. 
\begin{itemize}
	\item 
	The point $(x, y) = (1, 0)$ corresponds to the neutral global minimum if and only if
\begin{eqnarray}
	\lambda_2 > 0\quad \mbox{and}\quad \lambda_3 > 0\,.\label{conditions-BC} 
\end{eqnarray}
Following \cite{Ivanov:2014doa}, we label a generic neutral vev alignment as $(v_1, v_2, v_3)$ up to an overall factor:
\begin{eqnarray}
&&	\mbox{alignment $B$:}\quad B_1 = (1,\,0,\,0)\,, \quad B_2 = (0,\,1,\,0),\, \quad B_3 = (0,\,0,\,1)\label{points-B}\\
&&\mbox{alignment $C$:}\quad C_1 = (1,\,1,\,1)\,,\quad C_2 = (1,\,\omega,\,\omega^2)\,,\quad C_3 = (1,\,\omega^2,\,\omega)\label{points-C}\,.
\end{eqnarray}
The two sets of minima $B$ and $C$ correspond to the same point in the orbit space of the $\Sigma(36)$ 3HDM,
but they represent distinct manifolds in the Higgs spaces and, therefore, different points $\vec r$
is the space of bilinears, see again Appendix~\ref{appendix:bilinear}.
Other similar configurations such as $(\omega,\,\omega^2,\, 1)$ can be 
reduced to the ones already listed by an overall phase rotation of the three doublets;
for example, $(\omega,\,\omega^2,\, 1) = \omega (1,\,\omega,\,\omega^2)$ corresponds to the alignment $C_2$.
Note that all six minima $B$ and $C$ form a single $\Sigma(36)$ orbit; that is,
starting form any of the six minima and applying all the transformations 
from the $\Sigma(36)$ group, one can reach all the remaining minima.

	\item 
The point $(x, y) = (1, 1)$ corresponds to the neutral global minimum if and only if
\begin{eqnarray}
	\lambda_2 - 3 \lambda_3 > 0\quad \mbox{and}\quad \lambda_3 < 0\,.\label{conditions-AA'} 
\end{eqnarray}
The corresponding vev alignments are
\begin{eqnarray}
	&&\mbox{alignment $A$:}\quad A_1 = (\omega,\,1,\,1)\,, \quad A_2 = (1,\,\omega,\,1),\, \quad A_3 = (1,\,1,\,\omega)\label{points-A}\\
	&& \mbox{alignment $A'$:}\quad A'_1 = (\omega^2,\,1,\,1)\,, \quad A'_2 = (1,\,\omega^2,\,1),\, \quad A'_3 = (1,\,1,\,\omega^2)\label{points-Ap}
\end{eqnarray}
Notice that the relative phases between the vevs are not sensitive to the exact numerical values of the coefficients.
It was this phase rigidity (also called calculable phases) 
that drove the idea of geometric $CP$ violation in the $\Delta(54)$ 3HDM back in 1984 \cite{Branco:1983tn}.
We mention in passing that this model was analyzed in more detail in \cite{deMedeirosVarzielas:2011zw, Varzielas:2012nn, Ivanov:2013nla} and shown to be compatible with viable Yukawa sectors \cite{Bhattacharyya:2012pi, Varzielas:2013sla, Varzielas:2013eta}. 
We remind the reader that the 3HDM scalar sector invariant under the symmetry group $\Sigma(36)$ does not allow one to implement any form of 
$CP$ violation, neither explicit nor spontaneous \cite{Ivanov:2014doa}s.

	\item 
The point $(x, y) = (1/4, 0)$ corresponds to the charged breaking global minimum if and only if
\begin{equation}
	\lambda_2 < 0\quad \mbox{and}\quad \lambda_3 > 0\,.\label{conditions-G} 
\end{equation}
Just as before, there are six degenerate minima in the Higgs field space.
Two representative vev alignments, $G$ and $H$, are
\begin{equation}
	\lr{\phi_i} = v_G\left\{\doublet{0}{1}\,,\  \doublet{1}{0}\,,\  \doublet{0}{0}\right\}\,, \quad 
	v_H\left\{\doublet{0}{1}\,,\  \doublet{s_\alpha}{c_\alpha}\,,\  
	\doublet{-s_\alpha}{c_\alpha}\right\}\,.
\label{points-GH}
\end{equation}
Other minima can be obtained by applying the transformations from the group,
just as it was done for neutral minima.
	\item 
Finally, the point $(x, y) = (1/4, 1/4)$ corresponds to the charged breaking global minimum if and only if
\begin{equation}
	\lambda_2 - 3 \lambda_3 < 0\quad \mbox{and}\quad \lambda_3 < 0\,.\label{conditions-F} 
\end{equation}
Two representative vev alignments, $F$ anfd $F'$, can be chosen as
\begin{equation}
	\lr{\phi_i} = v_F\left\{\doublet{0}{\omega}\,,\  \doublet{s_\alpha}{c_\alpha}\,,\  
	\doublet{-s_\alpha}{c_\alpha}\right\}\,, \quad
	v_F\left\{\doublet{0}{\omega^2}\,,\  \doublet{s_\alpha}{c_\alpha}\,,\  
\doublet{-s_\alpha}{c_\alpha}\right\}\,, \label{points-FF'}
\end{equation}
where $c_\alpha = \cos\alpha = \cos(2\pi/3)$ and $s_\alpha = \sin\alpha = \sin(2\pi/3)$.
\end{itemize}
Since the three Higgs doublets transform under $\Sigma(36)$ as an irreducible triplet,
selecting any non-zero minimum will unavoidably break some of the symmetries.
However, none of the minima breaks $\Sigma(36)$ completely~\cite{Ivanov:2014doa}.
For each minimum point, there remains the little group of residual symmetries,
which are the transformations that leave the chosen vev alignment unchanged.
%In Appendix~\ref{appendix:residual} we list the residual symmetries for all the minima
%and discuss them for other extrema.

Further insights into the structural properties of the model, including the vev alignments, symmetry and $CP$ properties, 
can be gained if one pays attention not only to the transformations from the symmetry group $G$
but also to the transformations from $SU(3)$ which leave $G$ invariant, 
or ``symmetries of symmetries'' in the language of \cite{Fallbacher:2015rea}.
The potential remains form-invariant under such transformations, only up to reparametrization of coefficients,
which may provide additional links between different regimes of the same model.

%We end this section with a simple but useful remark on computing the depth at extremal points.
%Suppose we found a vev alignment $\lr{\phi_i}$ at which the potential has an extremum 
%of any sort,\footnote{We exclude a local maximum since, for a BFB potential, 
%it can exist only at the origin.} be it a minimum or a saddle point.
%Then, in the spirit of the virial theorem, the values of the quadratic 
%$V_2 \equiv V_2(\lr{\phi_i})$ and quartic $V_4\equiv V_4(\lr{\phi_i})$ parts 
%of the potential calculated at this point are linked:
%$V_2 = - 2 V_4 < 0$. Therefore, the depth of the potential can be evaluated only from its quadratic part:
%\begin{equation}
%	V(\lr{\phi_i}) = V_2 + V_4 = - \frac{1}{2}|V_2|\,.
%\label{virial}		
%\end{equation}
%Since the quadratic part of our model is very simple, one can use it to compute 
%the depth of the potential at any stationary point.

%%%%%%%%%%%%%%%%%%%%%

\section{Extrema of the $\Sigma(36)$ 3HDM}\label{section:numerical}

\subsection{The numerical procedure}\label{subsection:numerical-procedure}

In the previous Section, we listed all vev configurations the global minimum 
of the $\Sigma(36)$ 3HDM can have.
However, we aim to track all the extrema, 
not only the global minima. It turns out that {\tt Mathematica} cannot analytically solve 
the full extremization problem for the $\Sigma(36)$ 3HDM potential.
It becomes even more challenging for the case of the softly broken symmetry,
and although useful algebraic and geometric methods have been proposed 
in \cite{deMedeirosVarzielas:2021zqs,deMedeirosVarzielas:2022fnn}, they have limited reach and
cannot deal with the model with a generic set of soft symmetry breaking terms.

These observations force us to resort to numerical methods.
For this task, we need to choose numerical values of the coefficients.
Let us stress again that, in this paper, we explore structural properties of this toy model 
and do not perform any phenomenological fit. 
To this end, we set $\lambda_1 = 1$ and express all dimensional quantities in units of $m^2$;
for example, the potential depth will be plotted as the reduced depths $V/m^4$.
Thus we only need to explore the dependence on $\lambda_2$, $\lambda_3$
and, later, on the soft breaking parameters expressed in units of $m^2$.

Once the coefficients of the potential are fixed,
we find the extrema using the following numerical procedure.
We parametrize the Higgs doublets via 12 real components, 
compute the gradient of the potential, and numerically solve the system of equations $\vec \nabla V = 0$ 
starting from a random seed point.
The solver finds an extremum, checks its neutral/charge-breaking nature, finds the eigenvalues of the Hessian, 
and stores this information.
%which we convert in the vector of bilinears $(r_0, r_a)$, see Appendix~\ref{appendix:bilinear}, and store it. 

We repeat this procedure $N$ times for the same potential, each time starting from a different seed point,
and compare the new extremum with the ones already stored.
The value of $N$ must be sufficiently large so that new extrema no longer appear. 
We found empirically that, for a typical set of the coefficients, we need up to 
$N = 10,\!000$ seed points to detect all the extrema. 
This large number of seed points looks unavoidable; limiting the search to one thousand seed points
would often miss a few extrema.
%We believe that this technical observation is of importance to the community, 
%because most works run numerical minimization procedures only up to a few hundred times.

More details on the our numerical procedure can be found in Appendix~\ref{appendix:dependence}.

\subsection{3HDM with the exact $\Sigma(36)$}\label{subsection:extrema}

\begin{figure}[!htb]
	\centering
	\includegraphics[width=0.48\textwidth]{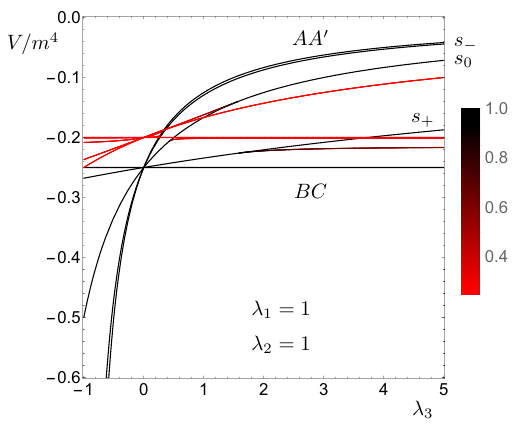}
%	\qquad 
	\includegraphics[width=0.48\textwidth]{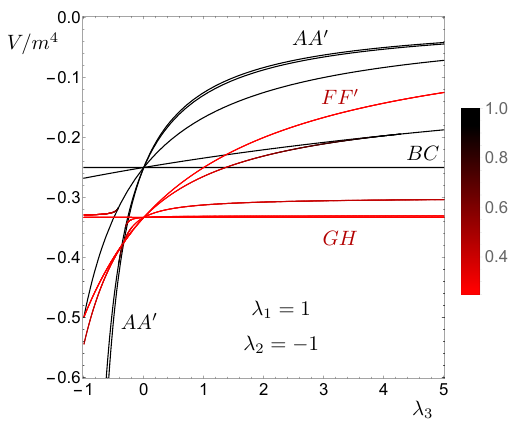}
	\caption{The dimensionless reduced depths of all the extrema of the $\Sigma(36)$ 3HDM potentia as a function of $\lambda_3$, computed for $\lambda_1 = 1$, $\lambda_2 = 1$ (left) and $\lambda_1 = 1$, $\lambda_2 = -1$ (right). 
	The color encodes the value of $x$, with the neutral extrema corresponding to $x=1$ (black)
	and the maximally charge-breaking extrema corresponding to $x=1/4$ (red).}
	\label{fig-scan-1}
\end{figure}

With the procedure described, let us now present results on the number and properties of all the extrema 
of the $\Sigma(36)$ 3HDM potential given in Eq.~\eqref{Vexact}.
To begin with, let us fix the parameter $\lambda_1 = 1$, choose $\lambda_2 = 1$ or $\lambda_2 = -1$, 
%\begin{equation}
%\lambda_1 = 1\,, \quad \lambda_2 = 1\,, \label{exact-point-1}
%\end{equation}
and scan over the range of $\lambda_3$ from $-1$ to $5$. The lower bound comes from the BFB conditions \eqref{BFB},
while the upper bound is chosen to include all the $\lambda_3$-dependent structural effects taking place in the scalar sector. 

The numerical results of this scan are shown in Fig.~\ref{fig-scan-1} for $\lambda_2 = 1$ (left pane) 
and $\lambda_2 = -1$ (right pane). 
We show here the reduced depths of the potential $V/m^4$ computed for each extremum.
The color indicates whether an extremum is neutral ($x=1$, shown in black)
or charge-breaking ($1/4 \le x<1$, shown in shades of red). Note that when a charge breaking
extremum merges with a neutral one, its color gradually shifts from bright red to black.
These plots demonstrate a rich spectrum of features, which are discussed in detail in Appendix~\ref{appendix:dependence}.
Here we briefly summarize the main observations and discuss how they affect phase transitions.

First, the $\Sigma(36)$ potential always has many extrema, either neutral or charge breaking. 
As the parameters change, the depths of these points evolve, and these extrema they can merge or split.
Depending on the value of $\lambda_3$, we observed 69 or 78 extrema in total, out of which six correspond to the global minima.
We remind the reader that, within the general 2HDM, the potential can admit at most two neutral minima 
plus four neutral and one charge-breaking saddle points, not counting the extremum at the origin.
Thus, the total number of non-trivial extrema in any 2HDM is at most seven.
The large number of extrema in the 3HDM comes from the higher dimensionality of the Higgs fields space.
%is due to the large discrete symmetry group we have which is partially broken at each extremum. 

The second observation is that the vev alignments for some extrema are rigid, 
in the sense that their vevs are calculable 
and stable upon smooth variation of $\lambda$'s, 
while the other extrema are ``floating''.
Among the rigid alignments, we have the global minimum, which for a positive $\lambda_2$ is attained 
at the points $A$ and $A'$ for $-1 < \lambda_3 < 0$
and at the points $B$ and $C$ for $\lambda_3 > 0$, 
while for a negative $\lambda_2$, it can be either charge breaking or neutral, 
in accordance with the results of the geometric minimization procedure.
It is also remarkable that all the special points mentioned in the previous section, even when they are not the global minima, 
are still present as extrema of the potential for any choice of the parameters. In addition, we found
other neutral rigid points which were not identified through the geometric procedure 
because they never corresponded to a global minimum. A detailed discussion
of the nature and regularities of these points is delegated to Appendix~\ref{appendix:dependence}.

Apart from the rigid points, we also observe in Fig.~\ref{fig-scan-1} other extrema, 
whose vev alignments and depths change smoothly as we vary $\lambda_3$.
All of these points, which we call the ``floating extrema'', lie in the charge-breaking part of the orbit space. 
Their ``charge breaking depth'', quantified by the parameter $1/4 \le x \le 1$ 
and shown in Fig.~\ref{fig-scan-1} with the shades of red color, varies smoothly as a function of $\lambda_3$.
%As we see in these plots, some extrema merge or split at certain values of $\lambda_3$, 
%and this process always involves a rigid extremum.

We numerically confirmed that the 3HDM potential with the exact $\Sigma(36)$ symmetry never develops 
a local minimum. We only have six degenerate global minima and saddle points.
This allows us to conclude that the exact $\Sigma(36)$ symmetry forbids first order phase transitions,
at least at the tree level. The only phase transitions which can take place are associated with passage through 
points of continuous symmetries.
In particular, for $\lambda_3 = 0$, the full symmetry group of the potential is promoted to $SU(3)$, 
instead of the finite group $\Sigma(36)$.
For negative $\lambda_2$, another transition takes place at $\lambda_3 = \lambda_2/3 = -1/3$. It is also triggered 
by an accidental continuous symmetry because, for these values of the parameters, 
the phase-sensitive terms of the Higgs potential \eqref{Vexact} 
can be grouped into a single combination $|\phi_1^\dagger \phi_2 + \phi_2^\dagger \phi_3 + \phi_3^\dagger \phi_1|^2$.
Notice that both transitions display a large finite jump in vevs, 
but since the coexisting minima lie on a manifold of degenerate vacua rather than separated by a finite barrier, 
these transitions are not of the first order. 

%\section{Softly broken $\Sigma(36)$ 3HDM}\label{section:numerical-softly}

\subsection{Introducing soft breaking terms}

Let us now explore how the structural features of the model change if we allow for soft breaking of $\Sigma(36)$.
The general quadratic terms can be written as 
$V_{\rm soft} = m^2_{ij} \phi_i^\dagger \phi_j$
with the hermitian matrix $m^2_{ij}$.
This matrix contains nine real parameters, including the three real $m_{ii}^2$ on the diagonal.
A generic set of soft breaking parameters would shift the positions and depths of all the extrema, 
lift symmetry-induced degeneracy, and perhaps change the number of minima and stationary points.
Exploring these changes in the entire 9-dimensional soft breaking parameter space and visualizing the results 
is a challenging task and, most likely, it will not be very revealing.
Besides, not all parameters play equal roles:
some of them may trigger structural changes,
while others only shift the numerical values of the observables.

Since our goal is to gain useful insights into evolution of the potential for various choices 
of these soft breaking parameters, we will not perform blind scans of the full 9-dimensional parameter space, 
but consider only such terms which
preserve the $\Z_3$ subgroup generated by the phase rotations $a = \mbox{diag}(1, \omega, \omega^2)$:
\begin{eqnarray}
	V_{\rm soft} = m_{11}^2 \phi_1^\dagger\phi_1 + m_{22}^2 \phi_2^\dagger\phi_2 + m_{33}^2 \phi_3^\dagger\phi_3\,.
	\label{soft3}
\end{eqnarray}
It is convenient to keep the overall mass scale $m^2$ unchanged, which implies 
$m_{11}^2  + m_{22}^2  + m_{33}^2  = 0$, leaving us with two independent 
soft breaking coefficients. We choose as free parameters the dimensionless quantities $\mu_1$ and $\mu_2$, 
which parametrize the plane in the $(m_{11}^2, m_{22}^2, m_{33}^2)$ space orthogonal to the $(1,1,1)$ direction.
The three coefficients $m_{ii}^2$ can then be expressed in the following way:
\begin{equation}
	\frac{m_{11}^2}{m^2} = \frac{\mu_1}{\sqrt{2}} - \frac{\mu_2}{\sqrt{6}}\,, \quad
	\frac{m_{22}^2}{m^2} = -\frac{\mu_1}{\sqrt{2}} - \frac{\mu_2}{\sqrt{6}}\,, \quad
	\frac{m_{33}^2}{m^2} = \frac{2\mu_2}{\sqrt{6}}\,.
	\label{soft-Z3}
\end{equation}
As we will see later, this parametrization will make obvious the third-order symmetry of the $(\mu_1,\mu_2)$ phase diagram
induced by the broken generator $b$ in Eq.~\eqref{Sigma36-generators}.

Let us also mention that one can organize soft breaking terms according to a different
principle, proposed and developed in \cite{deMedeirosVarzielas:2022fnn,deMedeirosVarzielas:2022kbj}
and recently extended to multi-flavon models \cite{Hagedorn:2023mrg}. 
Namely, one can introduce soft breaking terms which preserve the vev alignment of at least one minimum 
of the fully symmetric model. As shown in \cite{deMedeirosVarzielas:2022fnn}, this can be achieved 
if the chosen vev is an eigenvector of the matrix $m^2_{ij}$.
Our choice of the soft breaking terms satisfies this condition, as 
all the minima of type $B_i$ are the eigenvectors of the matrix $m^2_{ij}$.
However, one can in principle work with terms which softly break all the symmetries
of the original model and still preserve one of the minimum;
construction of such matrices was described in \cite{deMedeirosVarzielas:2022fnn} 
for the same $\Sigma(36)$ example as here. 
We delegate this more generic departure from the original symmetry to a future work.

\subsection{The phase diagram and the evolution of extrema in softly broken $\Sigma(36)$ model}\label{subsection-soft}

Let us now explore the soft breaking phase diagram and track possible sequences of phase transitions 
as we move across it.
Since our matrix $m^2_{ij}$ is already diagonal, all three vev alignments of type $B$ in Eq.~\eqref{points-B} 
are its eigenvectors and, according to \cite{deMedeirosVarzielas:2022fnn},
their alignments are preserved. The values of $v$ will, however, be different 
for $B_1$, $B_2$, and $B_3$.
The degeneracy among $B_i$ is lifted, and we now have three isolated local minima at different depths,
separated by barriers. In contrast, the three points $C$ in Eq.~\eqref{points-C}
remain at the same depth because they are linked by the unbroken symmetry $a$.
We would like to stress here that the vev alignments $C_i$ are not the eigenvectors of the soft breaking matrix
and they do evolve. By labeling an extremum as, for example, $C_1$, we refer to the point which
would correspond to the vev alignment $(1,1,1)$ in the limit of vanishing soft breaking terms.

\begin{figure}[!htb]
	\centering
	\includegraphics[width=0.5\textwidth]{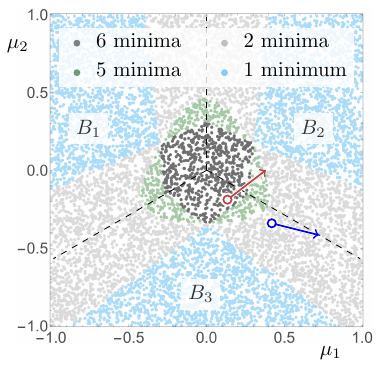}
	%	\qquad 
	%\includegraphics[width=0.48\textwidth]{soft-115-zoomed.pdf}
	\caption{The phase diagram of the softly broken $\Sigma(36)$ model in the plane of dimensionless 
		soft breaking parameters $(\mu_1, \mu_2)$ for $\lambda_1 = 1$, $\lambda_2 = 1$, $\lambda_3 = 5$. 
	The color encodes the total number of minima; when the minimum is unique, it is explicitly labeled.
	The two circles with arrows correspond to the thermal evolution of the two benchmark models 
	discussed in Section~\ref{subsection:benchmark}: benchmark model 1 (blue) and benchmark model 2 (red).}
	\label{fig-scan-soft-1}
\end{figure}

In Fig.~\ref{fig-scan-soft-1} we present the results of a scan in the plane of the dimensionless soft breaking parameters $(\mu_1, \mu_2)$ 
for the choice of the quartic coefficients $\lambda_1 = 1$, $\lambda_2 = 1$, $\lambda_3 = 5$. 
The plane is composed of three ``zones of influence'', separated by dashed lines, 
each zone corresponding to the deepest minimum being
$B_1$, $B_2$, or $B_3$, as labeled on the plot.
The evident three-fold symmetry of the phase diagram is due to the fact that, 
upon the cyclic permutation $\phi_1 \mapsto \phi_2 \mapsto \phi_3 \mapsto \phi_1$, 
which induces the $2\pi/3$ rotation of the diagram, 
the physical picture does not change, up to the permutation of the minima.

Regions with different colors correspond to models with different total numbers of minima.
The central dark gray region corresponds to the potentials which still possesses 
all six minima $B_i$ and $C_i$, which were present for the exact $\Sigma(36)$ symmetry, although they are now located 
at different depths.
As we move on the $(\mu_1, \mu_2)$ plane away from the origin, we cross a boundary, at which one of the $B_i$ minima
merges with saddle points and becomes itself a saddle point.
Thus, in the olive region, we have only five minima.
Upon further growth of the soft breaking coefficients, three points $C_i$ cease to be the minima 
and, merging with low lying saddle points, become saddle points themselves. 
With a further increase of the soft breaking parameters, one can either stay in the light gray region, 
with two minima at different depths, or enter the light blue region where only one minimum survives.
We checked that the boundaries at which $B_i$ successively disappear are given by straight lines,
while the region where $C_i$ remain the minima is bounded by a non-trivial cycloid-like curve.

In the above diagram, we described the regions where points $B_i$ or $C_i$ remain (local) minima.
When we cross their boundaries, these extrema do not disappear but just merge with other
saddle points and become saddle points themselves. Since the nature of the deepest saddle point 
is very relevant for the phase transition dynamics, we now would like to check
the evolution of all the extrema as we vary $\mu_1$ or $\mu_2$. 

\begin{figure}[!htb]
	\centering
	\includegraphics[width=0.48\textwidth]{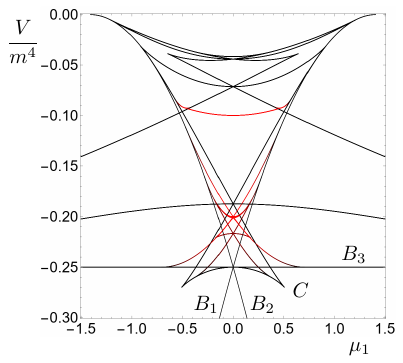}
	%	\qquad 
	\includegraphics[width=0.48\textwidth]{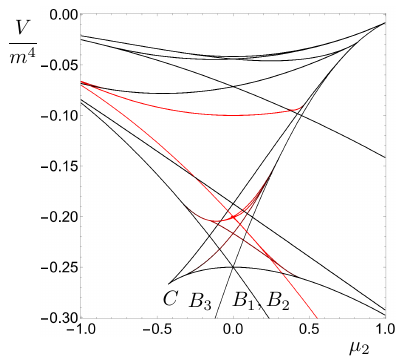}
	\caption{The evolution of the reduced depths of all the extrema for the model
		with $\lambda_1 = 1$, $\lambda_2 = 1$, $\lambda_3 = 5$ 
		as a function of the dimensionless 
		soft breaking parameters $\mu_1$ calculated at $\mu_2 = 0$ (left)
		and as a function of $\mu_2$ taken at $\mu_1 = 0$ (right).
	Black lines correspond to neutral extrema, red lines show charge breaking extrema,
	with the intensity of color denoting the charge breaking ``depth''.}
	\label{fig-scan-soft-2}
\end{figure}

In Fig.~\ref{fig-scan-soft-2}, we show how the depths of all the extrema change as
we scan the phase diagram along $\mu_1$ keeping at $\mu_2 = 0$ (left plot)
and along $\mu_2$ keeping $\mu_1 = 0$ (right plot).
Starting from the $\Sigma(36)$-symmetric case and adding positive $\mu_1$, 
we observe the point $B_2$ to become the unique global minimum, 
while points $B_1$, $B_3$, and $C$ to stay local minima. 
At the same time, the deepest saddle points, too, spread into several non-equivalent saddle points,
the deepest among them approaching the minimum $C$.
They merge around $\mu_1 \approx 0.4$, at the edge of the olive region in Fig.~\ref{fig-scan-soft-1}.
Above this value, $C$ is no longer a minimum but continues as a saddle point
until about $\mu_1 \approx 0.5$, where it disappears as an extremum
and just remains an inflection-type feature on the slope of the potential.
Note that point $B_1$ remains a minimum until about $\mu_1 \approx 0.4$,
even though it becomes a shallow minimum which lies higher than
several charge breaking saddle points.

The right plot of Fig.~\ref{fig-scan-soft-2} shows the $\mu_2$ evolution of the extrema at $\mu_1=0$;
this trajectory passes vertically through the plot in Fig.~\ref{fig-scan-soft-1}.
For this choice of soft breaking parameters, points $B_1$ and $B_2$ remain degenerate
and, for positive $\mu_2$, they are the global minima.
Note that, starting from $\mu_2 \approx 0.3$, the deepest saddle point 
is a different charge breaking point, the one shown in \eqref{points-GH}.

In general, we see that as we move from the exactly symmetric case, many previously degenerate extrema split into families of 
extrema at different depths. This includes, in particular, several coexisting minima,
separated by barriers of different heights, so that a non-trivial sequence of first-order phase transitions
can be anticipated. As the coefficients in the soft breaking coefficients grow further, many saddle points merge and then disappear,
so that the structure of stationary points simplifies.
This is mainly driven by the coefficients in front of some $\phi_i^\dagger \phi_i$ becoming positive.

Let us also briefly discuss the case of negative $\lambda_3$, for which the global minimum
of the symmetric model corresponds to points $A,A'$.
Upon adding the soft breaking terms of Eq.~\eqref{soft-Z3}, 
the six points $A,A'$ split into a multiplet of individual minima.
Although the exact vev alignments given in Eqs.~\eqref{points-A},~\eqref{points-Ap} is not preserved, 
we still can label them $A_i, A_i'$
according to their fully symmetric limit.
Since these soft breaking terms are $CP$ invariant, 
the minima of type $A$ and $A'$ are pairwise degenerate.
However each of these minima is now $CP$ violating;
thus, we observe spontaneous $CP$ violation driven by soft breaking terms.

We also produced the phase diagram, in which, just as in the previous example, 
we observed several regions corresponding to different number of minima. 
We studied the evolution of extrema, with an equally rich
list of options for phase transitions.
A new feature which arises in this case is that,
in large soft breaking parameters, all minima of type $A$ and $A'$ disappear
and the global minimum is again attained at one of the points $B$.
Thus, the blue regions of Fig.~\ref{fig-scan-soft-1} persist even for negative $\lambda_3$.

\subsection{Charge breaking minima and phase transitions}

\begin{figure}[!htb]
	\centering
	\includegraphics[width=0.45\textwidth]{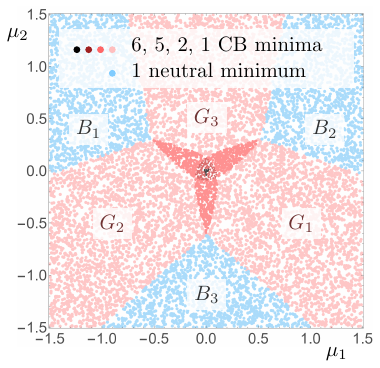}
	\qquad 
	\includegraphics[width=0.48\textwidth]{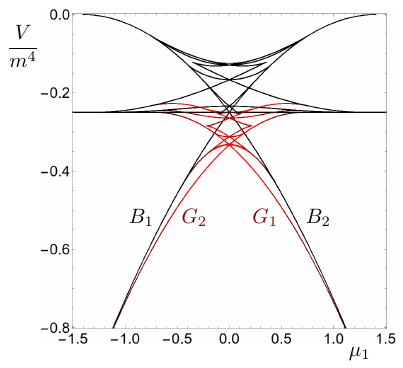}
	\caption{Left: the phase diagram of the softly broken $\Sigma(36)$ model in the plane of the dimensionless 
		soft breaking parameters $(\mu_1, \mu_2)$ 
		for $\lambda_1 = 1$, $\lambda_2 = -1$, $\lambda_3 = 1$. Right:
	the $\mu_1$ evolution, at $\mu_2=0$, of the extrema in this model.}
	\label{fig-scan-soft-3}
\end{figure}

Let us now turn to the scenario, 
in which the model with an exact $\Sigma(36)$ symmetry exhibits a charge breaking global minimum.
In Fig.~\ref{fig-scan-soft-3}, left, we show the phase diagram for the model 
with $\lambda_1 = 1$, $\lambda_2 = -1$, $\lambda_3 = 1$.
Just as before, the central region, corresponding to small soft breaking terms, 
contains several charge breaking minima, which gradually disappear as the soft breaking terms become stronger.
Away from the central region, the situation simplifies, with only one minimum remaining.
It is interesting, however, that depending on the proportion between $\mu_1$ and $\mu_2$,
this minimum can be neutral, of type $B$, or charge breaking, of type $G$.
For example, at a large positive $\mu_2$, the minimum can be at
$B_1 \propto (1, 0, 0)$,  $B_2 \propto (0, 1, 0)$,
or the charge-breaking alignment
\begin{equation}
	G_3 \propto \left\{\doublet{0}{\cos\psi}\,,\  \doublet{\sin\psi}{0}\,,\  \doublet{0}{0}\right\}\,.
	\label{G3}
\end{equation}
The values $\psi = 0$ and $\psi = \pi/2$ correspond to the points $B_1$ and $B_2$,
respectively, while for a generic $\psi$, the alignment $G_3$ smoothly interpolates between these neutral points.
This is also confirmed by Fig.~\ref{fig-scan-soft-3}, right, where we show the
evolution of all the extrema as a function of $\mu_1$ for fixed $\mu_2 = 0$.

We conclude that, as we shift across the phase diagram, we observe CB-to-neutral or neutral-to-CB phase transitions.
They are of the second order because, as it turn out, neutral and CB minima 
do not coexist in this example. They can become the first order beyond the tree level,
as it happened in the 2HDM when the one-loop effective potential treatment of \cite{Aoki:2023lbz}
improved over the tree-level approach of \cite{Ginzburg:2009dp,Ivanov:2008er}.
However we do observe the first order phase transitions inside the CB phase even at the tree level, 
just because several CB minima can certainly coexist at small soft breaking parameters.

A general take away message is that all these structural opportunities, 
with their possibly intriguing phase transitions dynamics,
arise thanks to soft breaking terms.

\section{Charge breaking bubble wall between two neutral minima}\label{section:domain}

\subsection{Charge breaking bubble wall in the 2HDM}\label{subsection:2HDM}

As we observed in Section~\ref{section:numerical}, the $\Sigma(36)$ 3HDM potential with $\lambda_2 > 0$ 
and a sufficiently large $\lambda_3$ contains the deepest saddle point which is charge breaking, not neutral.
Before exploring it, we would like to remark that such a situation does not require three Higgs doublets
and is possible within the 2HDM. 
This simple fact remained for a long time unexplored 
and, to our best knowledge, it was first explicitly pointed out only in the very recent
publication \cite{Sassi:2023cqp}.
To give a concrete example tractable analytically, let us consider the 2HDM potential
\begin{eqnarray}
	V_{\rm \tiny 2HDM} &=& - m^2\left(\phi_1^\dagger\phi_1 + \phi_2^\dagger\phi_2\right)
	+ \frac{\lambda_1}{2} \left[(\phi_1^\dagger\phi_1)^2 + (\phi_2^\dagger\phi_2)^2\right]\nonumber\\
	&& + \ \lambda_3(\phi_1^\dagger\phi_1)(\phi_2^\dagger\phi_2)+ \lambda_4 |\phi_1^\dagger \phi_2|^2 + \frac{\lambda_5}{2} \left[(\phi_1^\dagger\phi_2)^2 + (\phi_2^\dagger\phi_1)^2\right]\,.\label{V-2HDM}
\end{eqnarray}
It is invariant under the sign flips of each doublet, $\phi_1 \leftrightarrow \phi_2$, and the usual $CP$ transformation,
which combine, group theoretically, into the $(\Z_2)^3$ group acting in the space of bilinears \cite{Ivanov:2005hg,Nishi:2006tg}.
%This potential can also be obtained by imposing a single generalized $CP$ symmetry: CP$_g^{(i)}$ in the notation of \cite{Maniatis:2007vn,Maniatis:2007de} or CP2 in the notation of \cite{Ferreira:2010yh}. 
This potential can also be obtained by imposing a single generalized $CP$ symmetry 
\cite{Maniatis:2007vn,Maniatis:2007de,Ferreira:2010yh}. 
Let us also assume that the quartic parameters $\lambda_i$ satisfy, in addition to the usual BFB conditions,
the following inequalities: $\lambda_3 > \lambda_1$ and $\lambda_4 - |\lambda_5| > 0$.
Then, straightforward algebra allows us to find all the extrema of this potential. 
We get a pair of degenerate global minima at
\begin{equation}
	\sqrt{2}(\lr{\phi_1^0}, \lr{\phi_2^0}) = (v, 0)\  \mbox{or} \  (0, v)\,, 
	\ \mbox{where}\ v^2 = \frac{2m^2}{\lambda_1}\,,
	\quad V = - \frac{m^4}{2\lambda_1}\,,
\end{equation} 
two pairs of neutral saddle points at
\begin{eqnarray}
	\sqrt{2}(\lr{\phi_1^0}, \lr{\phi_2^0}) &=& (v', \pm v')\,, \ 
	\mbox{with}\ v^{\prime 2} = \frac{2m^2}{\lambda_1 + \lambda_3 + \lambda_4 + \lambda_5}\,,
\quad V = - \frac{m^4}{\lambda_1 + \lambda_3 + \lambda_4 + \lambda_5}\,,\nonumber\\
	\sqrt{2}(\lr{\phi_1^0}, \lr{\phi_2^0}) &=& (v'', \pm iv'')\,, \ 
\mbox{with}\ v^{\prime\prime 2} = \frac{2m^2}{\lambda_1 + \lambda_3 + \lambda_4 - \lambda_5}\,,
\quad V = - \frac{m^4}{\lambda_1 + \lambda_3 + \lambda_4 - \lambda_5}\,,
\end{eqnarray}
as well as a charge breaking saddle point at
\begin{equation}
	\lr{\phi_1^0} = \frac{1}{\sqrt{2}}\doublet{0}{u}\,, \  
	\lr{\phi_2^0} = \frac{1}{\sqrt{2}}\doublet{u}{0}\,, \ \mbox{where}\ u^2 = \frac{2m^2}{\lambda_1 + \lambda_3}\,,
	\quad V = - \frac{m^4}{\lambda_1 + \lambda_3}\,.	
\end{equation}
With the parameters satisfying $\lambda_3 > \lambda_1$ and $\lambda_4 - |\lambda_5| > 0$, 
the charge breaking extremum becomes the deepest saddle point.

By adding soft breaking terms, we can avoid degeneracy between the two minima while keeping the CB saddle point
the deepest one. If the early Universe temporarily resided in the local metastable minimum, it could
decay, via tunneling or thermal fluctuations, to the true minimum, 
see for example the pedagogical expositions in \cite{Rubakov:2002fi,Hindmarsh:2020hop}.
As a result, bubbles of true vacuum will spontaneously form and expand in the false vacuum background,
sweeping across the primordial hot plasma and eventually colliding and merging.
Not aiming at any accurate computation of the nucleation rate, 
%(the probability of bubble appearance per unit volume and unit time), 
we only mention that, neglecting thermal fluctuations and treating the tunneling integral in the thin-wall approximation,
we find the exponentially suppressed value $\exp(- S_E)$, 
where $S_E$ is the euclidean action calculated for the so-called bounce solution \cite{Coleman:1977py}.

\begin{figure}[!htb]
	\centering
	\includegraphics[width=0.4\textwidth]{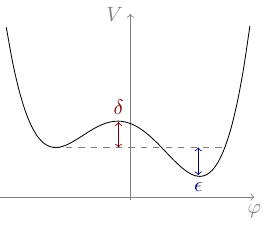}
	\caption{A schematic profile of a titled double-well potential.}
	\label{fig-tilted}
\end{figure}

In the simplest example of one real scalar field with a tilted double-well self-interaction 
potential with the quartic coupling $\lambda$ as schematically depicted in Fig.~\ref{fig-tilted},
$S_E$ depends on the height of the barrier $\delta$ and 
the difference $\epsilon$ between the depths of the two minima,
in the following way:
\begin{equation}
	S_E = C \left(\frac{\delta}{\epsilon}\right)^3\cdot \frac{1}{\lambda}\,.\label{SE-estimate}
\end{equation}
Here, $C$ is the numerical coefficient of the order of a few hundred, which is inessential for the present qualitative discussion.
A good rule of thumb is that, for reasonable values of $\lambda$, the values $\delta/\epsilon \ll 1$ 
lead to a very quick tunneling, while $\delta/\epsilon \gg 1$ are characteristic of extremely long-lived
trapped metastable states.

With these preliminary remarks, let us now consider a multi-field potential, with two different minima and several competing saddle points.
It may happen that several paths passing through different saddle points exist. 
For each competing path, the action involves the barrier height $\delta$ of its saddle point 
and an effective $\lambda$, which essentially depends on how long this optimal path is.
Although these paths are not necessarily straight,
we can roughly estimate $1/\lambda \sim (\Delta v)^4/\delta$, where $\Delta v$ 
is the straight segment distance between the minimum and the saddle point.
These heuristic arguments lead us to the general conclusion that, among several competing paths,
the minimal action is achieved along the path passing through the deepest saddle point,
provided it does not lie too far from the two minima it connects.
Applying this qualitative criterion to the 2HDM example given above, we see that 
the path passing through the charge breaking saddle point is indeed optimal.

Thus, the phase transition will proceed through formation, expansion, and coalescence
of bubbles of true vacuum with CB bubble walls.
It is interesting to see whether these CB bubble walls, sweeping through plasma, 
has any consequences for baryogenesis or leads to fermion charge separation, capable of seeding large-scale charge
inhomogeneities and leading to primordial magnetic fields.
The recent study \cite{Sassi:2023cqp} provided only a few first insights 
into possible phenomena which could take place during this disruptive process, 
and more studies are definitely needed.

\subsection{Charge breaking bubble walls in the $\Sigma(36)$ 3HDM}

Returning to the $\Sigma(36)$ 3HDM, let us first remark that we have identified two regimes
where neutral minima of different depths are separated by a charge breaking domain wall.
The first one exists already for the exactly symmetric case, while the second one
appears upon adding a significant soft breaking term.

First, when studying the phase diagram for the softly broken case, we selected $\lambda_3 = 5$
which guaranteed that the deepest saddle point was charge breaking,
see Fig.~\ref{fig-scan-1}. 
This is a floating point, so that its vev alignment smoothly depends
on the values of the coefficients, but
in the limit of large $\lambda_3$, it asymptotically approaches
\begin{equation}
	\lr{\phi_i} \propto 
	\left\{\doublet{0}{1}\,,\  \doublet{0}{1}\,,\  
	\doublet{\sqrt{3}}{1}\right\}\,.
	\label{deepest-saddle}
\end{equation}
This asymptotic vev alignment corresponds to the point $(x,y) = (1/2, 0)$.
Of course, by applying $\Sigma(36)$ transformations, we can find other vev configurations 
corresponding to this point; in total, there are nine.

Let us now add soft breaking terms with moderately large $\mu_1$ passing from negative to positive values, 
while keeping $\mu_2 = 0$. Initially, the global minimum is at $B_1$.
As we pass the symmetric point and move into the positive $\mu_1$ region,
$B_1$ becomes a local minimum ready to tunnel to a lower-lying minimum,
either $C$, $B_3$, or the global minimum $B_2$. 
To which minimum it will actually tunnel is not immediately clear
and requires a dedicated numerical study.
What is certain is that the saddle point through which 
this transition occurs is charge breaking, and the natural expectation
is that it is the point in Eq.~\eqref{deepest-saddle}.

If $\mu_1$ increases further, this deepest saddle point approaches the neutral minima of type $C$,
merges with them resulting in neutral saddle points, see again Fig.~\ref{fig-scan-soft-2}, left.
However if, in addition to the moderately large $\mu_1$, 
we add a sufficiently large positive $\mu_2$, we enter the second regime with CB saddle points, 
similar to Fig.~\ref{fig-scan-soft-2}, right.
This time, the two competing minima are of type $B_1$ and $B_2$,
while the CB saddle point is of type $G_3$, as in Eq.~\eqref{G3}.
This regime, which closely resembles the 2HDM example shown above, persists even for large $\mu_2$, 
and the deepest CB saddle point never merges with the neutral vev manifold.

\subsection{Benchmark models}\label{subsection:benchmark}

To illustrate the above general discussion, we provide in this subsection 
two benchmark models with softly broken $\Sigma(36)$, which lead to first order phase transitions 
involving charge breaking bubble walls.
In the first model the CB nature of the wall in unambiguous,
while in the second model we observe several competing saddle points of different nature,
which potentially leads to the remarkable situation of several bubble of the same new vacuum
separated by different bubble walls.

We begin by writing, at zero temperature, the quadratic part of the potential as
\begin{eqnarray}
	V_2 = \tilde m_{11}^2 \phi_1^\dagger\phi_1 + \tilde m_{22}^2 \phi_2^\dagger\phi_2 + \tilde m_{33}^2 \phi_3^\dagger\phi_3\,,
	\label{qadratic-T0}
\end{eqnarray}
where the coefficients $\tilde m_{ii}^2 \equiv - m^2 + m_{ii}^2$ include the overall $m^2$ 
from the fully symmetric potential Eq.~\eqref{Vexact} and the soft breaking terms of Eq.~\eqref{soft3}.
We adjust the values of $m^2$ and $\lambda_1$ to reproduce $v = 246$~GeV and $m_h = 125$~GeV.
As suggested by the analysis in the previous section, 
we often encounter minima of type $B_i$. Their vevs are the eigenvectors 
of the soft breaking terms we use, which leads to scalar alignment even in the softly broken case \cite{deMedeirosVarzielas:2021zqs}.
This allows us to fix the parameters $m^2 = m_h^2/2$ and $\lambda_1 = m^2/v^2 \approx 0.129$ at zero temperature.
Just as in Section~\ref{subsection-soft}, we also fix $\lambda_2 = \lambda_1$ and $\lambda_3 = 5\lambda_1$, 
so that we can rely on the phase diagram shown in Fig.~\ref{fig-scan-soft-1} 
and the extrema evolution in Fig.~\ref{fig-scan-soft-2}. 

The two benchmark models then differ only by the zero-temperature values of the soft-breaking parameters $\mu_1$ and $\mu_2$:
\begin{equation}
	\mbox{model 1:}\quad \mu_1 = 0.5\,, \ \mu_2 = -0.45\,, \qquad
	\mbox{model 2:}\quad \mu_1 = 0.2\,, \ \mu_2 = -0.2\,.
	\label{thermal-parameters}
\end{equation}
The two choices of $(\mu_1,\mu_2)$ are indicated in Fig.~\ref{fig-scan-soft-1}, by circles with arrows.
The first model corresponds to the point just below the right dashed line,
while the second model refers to the point near the boundary between the black and blue regions. 
In both cases, the zero-temperature minimum is at the vev alignment $B_3 = (0,0,1)$.

We now add temperature evolution.
At non-zero temperature, the shape of the potential changes, which can be analyzed 
within the one-loop thermal effective potential \cite{Dolan:1973qd,Cline:1996mga}.
To make the physical consequences transparent, let us focus on the main effect in the high-$T$ approximation,
the thermal $T^2$ contributions to the scalar masses.
These are induced by one loop corrections to scalar field propagators: 
\begin{equation}
	\tilde m_{ii}^2 \to \tilde m_{ii}^2(T) = \tilde m_{ii}^2 + c_i T^2\,.\label{thermal-1}
\end{equation} 
Once we know $\tilde m_{ii}^2$ and $c_i$, we can represent the thermal history of the scalar sector 
as a trajectory on the $(\mu_1,\mu_2)$ plane, see a 2HDM example in the recent study \cite{Aoki:2023lbz}.
These trajectories are straight, and their directions for the two benchmark models 
are shown by the arrows in the Fig.~\ref{fig-scan-soft-1}.

Due to the $\Sigma(36)$ symmetry of the quartic interactions, the scalar one-loop contributions are identical 
for all three Higgs doublets.
The gauge boson contributions are also identical and equal to the standard expression for electroweak doublets.
Thus the coefficients $c_i$ can only differ in how each doublet couples to fermions:
\begin{equation}
	c_i = \frac{7\lambda_1 + \lambda_2 + 2\lambda_3}{6} + \frac{3g^2 + g^{\prime 2}}{16} 
	+ \frac{1}{12}\sum_f N_c [y^{(f)}_i]^2\,.\label{thermal-2}
\end{equation} 
We explicitly indicated here the color factor $N_c=3$ for quarks and $N_c=1$ for leptons. 
The fermion loop contributions depend on the Yukawa structures and, 
in particular, on the top-quark coupling with the three Higgs doublets.
%Of course, construction of a full benchmark model requires building the Yukawa sector and checking whether 
%it correctly reproduces the fermion masses and mixing angles.
In both benchmark models, we assume that the top-quark couples only to the third Higgs doublet
with the SM-like coupling: $y^{(t)}_1 = y^{(t)}_2 = 0$, $y^{(t)}_3 =\sqrt{2}m_t/v \approx 1$. 
Of course, other Yukawa sector models are possible and should be included when
building a realistic $\Sigma(36)$ 3HDM model.

\begin{figure}[!htb]
	\centering
	\includegraphics[width=0.46\textwidth]{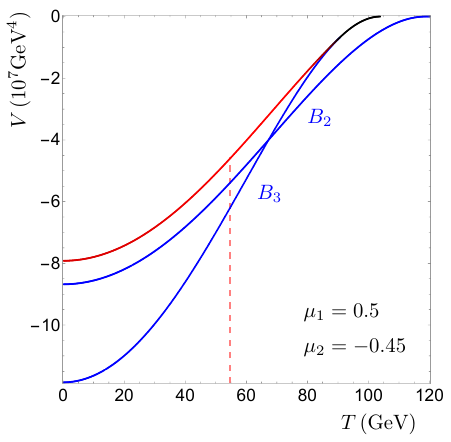}
	\qquad 
	\includegraphics[width=0.46\textwidth]{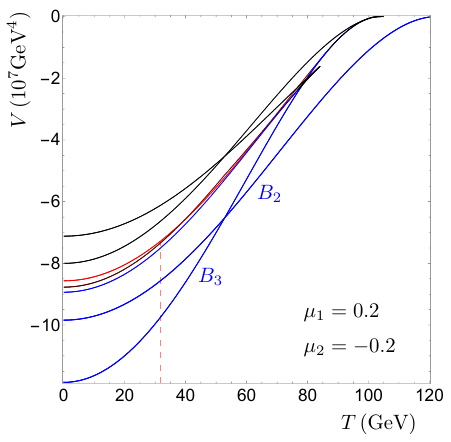}
	\caption{Temperature dependence of the depths of the minima (shown in blue) and the few deepest saddle points
		(neutral shown in black, charge-breaing shown in red) 
		of the Higgs potential in the two benchmark models, see the main text for the parameters.
	The vertical pink dashed lines show the temperatures when the barrier height due to the deepest saddle point is equal
to the difference between the depths of the two minima.}
	\label{fig-T-evolution}
\end{figure}

In Fig.~\ref{fig-T-evolution}, we show temperature evolution of the depths of the minima $B_2$ and $B_3$ as well as
the few lowest lying saddle points in the benchmark model 1 (left) and model 2 (right).
For model 1, we show only the deepest saddle point, %which is charge-breaking below $T \approx 90$ GeV;
as all additional saddle points lie significantly higher.
Let us track the evolution of the Higgs potential as the Universe cools down. 
At very high temperatures, the quadratic coefficients are all positive and the electroweak symmetry is restored, 
the minimum staying at $v = 0$. Around $T = 120$ GeV, the global minimum shifts to the alignment $B_2 = v(T)(0,1,0)$, 
with the value of $v(T)$ gradually increasing as the temperature drops.
Below $T \approx 90$ GeV, the vev alignment $B_3$ transforms from a saddle point to a minimum,
and its depth quickly grows. At 70 GeV, $B_3$ becomes the global minimum. However the Universe still resides
in the local minimum $B_2$ due to a significant potential barrier which separates the two minima.
This potential barrier goes over the deepest saddle point which is charge breaking and is shown 
in Fig.~\ref{fig-T-evolution} in red.

The actual phase transition happens at a lower temperature, when the difference of the depths of the two minima 
becomes comparable to the saddle point barrier height, or $\delta/\epsilon \sim 1$ in the notation of Section~\ref{subsection:2HDM}.
To guide the eye, the pink dashed line in Fig.~\ref{fig-T-evolution}, left, indicates 
the temperature when $\delta/\epsilon = 1$, which is around $55$~GeV.
Note that we do not claim that the phase transition occurs exactly at this point, 
for such a claim would require an exact numerical computation of the bounce integral
in the multi-Higgs space and the evaluation of the bubble nucleation rate.
We just point that it will happen at temperatures somewhat below the point of the $B_2/B_3$ crossing,
and that it will be a first order phase transition.
Since the deepest saddle point is charge breaking and all other saddle points lie significantly higher, 
this phase transition definitely proceeds via emergence 
of bubbles of the vacuum $B_3$ in the background of the false vacuum $B_2$,
separated by charge breaking bubble walls. 
This is exactly the regime predicted by our general analysis.

In the second benchmark model, shown in Fig.~\ref{fig-T-evolution}, right,
temperature evolution proceeds similarly. However, the choice of the zero-temperature 
point $(\mu_1,\mu_2)$ now leads to a much more intricate evolution of the saddle points.
We see that in the temperature region where a phase transition is expected to occur
we have two families of saddle points,
plus yet another family of local minima of type $C$.
We find that the two competing saddle points are charge breaking, 
although one of them is close to a neutral Higgs configuration, 
and are well separated in the Higgs field space.
These extrema display similar but not exactly identical temperature evolution.
Thus, within the benchmark model 2, our simplified analysis is unable to answer 
which of the two competing saddle points gives the higher nucleation probability.
It is, in fact, well possible that new vacuum regions with both types of bubble walls 
emerge, expand and merge.

The lesson we learn from benchmark model 2 is that situations with competing saddle points
are rather generic in the symmetry-based 3HDMs.
Our simplistic analysis only indicates that such peculiar regime indeed exists
but cannot accurately predict the dynamics of phase transition.
Quantitative understanding of the evolution of the Universe, 
as well as its implications for gravitational wave productions, baryogenesis, and primordial magnetic field generation,
requires a dedicated numerical study at the level of finite-temperature one-loop effective potential.

\section{Discussion and conclusions}\label{section:discussion}

In this work, we presented an exploratory study of a multi-Higgs potential with a large symmetry group, 
either exact or softly broken,
in which we not only looked at the global minimum but also tracked all extrema of the potential.
The calculations were done for $\Sigma(36)$ 3HDM, but we believe the lessons learned
can be useful in other multi-Higgs models.
Here are the main observations made in the course of this study. 
\begin{itemize}
	\item 
	Potentials with large finite symmetry groups, even with very few free parameters, 
	possess surprisingly rich structure of minima and saddle points.
	In the fully symmetric 3HDM case, we observed up to 78 extrema, either neutral and charge breaking,
	which is a significant increase with respect to the 2HDM, where the potential
	can possess at most seven stationary points.
Vev alignments of the extrema can be ``rigid'' (stable against smooth variation of the free parameters)
or ``floating'' (smoothly shifting as the parameters change).
There are also bifurcation points where saddle points merge or split, with interesting symmetry driven patterns.
%In the softly broken case, local minima can merge with saddle points to become saddle points themselves,
%while saddle points can merge and disappear altogether.
	\item 
	When we searched for all the stationary points via numerical methods, we were forced to use
	several thousand seed points in order not to miss any extremum. This was a significant bottleneck in our study,
	and it calls upon more efficient methods for locating all stationary points of a multi-Higgs potential.
	\item 
	There are regions in the parameter space where the deepest saddle point is charge breaking.
	As a result, there can exist charge-breaking domain walls separating neutral minima.
	We also gave an explicit 2HDM example with the same feature.
	\item
	In the softly broken case, we found two distinct regimes, in which tunneling between two neutral minima at different depths 
	proceeds via expansion and collision of charge-breaking bubble walls.
%	\item 
%	We also found regions in the parameter space where CB-to-neutral phase transitions become possible,
%	and where the first order CB-to-CB phase transitions take place.
\end{itemize}
In short, we find it remarkable that such a structurally simple model as $\Sigma(36)$ 3HDM with soft symmetry breaking
allows for several phase transition regimes without any fine-tuning.
We believe that a detailed numerical investigation of many of these processes may lead 
to new insights into possible dynamics of the early Universe.

A particularly intriguing phenomenon is tunneling between neutral vacua
which proceeds via expansion and collision of charge-breaking bubble walls.
Although this feature was recently identified in \cite{Sassi:2023cqp} within the 2HDM,
the multi-Higgs example we considered here offers novel opportunities. 
First, we observe two different regimes for this phenomenon, corresponding to weak 
and to strong soft breaking terms, with potentially different consequences.
Second, the 3HDM example possesses many more minima and saddle points than the 2HDM,
with several competing phase transitions and even competing tunneling paths
between the same pair of minima.

To illustrate some of these findings, we presented two benchmark models, 
including one in which saddle points of different nature indeed compete.
This situation can potentially lead to a remarkable cosmological scenario in which 
bubble of the same true vacuum appear and expand in the false vacuum background, 
but these bubbles possess bubble walls of different nature.

Numerous questions for follow-up studies naturally emerge.
Staying within the same toy model, the softly broken $\Sigma(36)$ 3HDM,
one can explore in more detail the exact sequence of phase transitions. 
In particular, for a potential with generic soft breaking terms, 
one needs to identify the exact trajectory in the Higgs field space
which minimizes the euclidean action between a chosen pair of minima.
As several saddle points can compete, it may happen that the bubble nucleation rate
could depend in a highly non-trivial way on the parameters of the potential.
Also, if the symmetry group is not fully broken, one can perhaps observe the situation when
different bubbles produce different true vacua at the same depth.
It will be interesting to see whether genuine stable domain walls can be produced 
or whether one of the competing vacua eventually takes over.
One can also imagine the situation with two phase transitions taking place in rapid succession,
so that the second one sets in before the bubbles from the first phase transition 
have already expanded and fully merged.
Extensive numerical simulations are required to verify all these peculiar evolution regimes. 

%Another question is to quantitatively estimate the bubble nucleation rate and, in this way, to find which part of the 
%$(\mu_1,\mu_2)$ plane can support the long-lived metastable vacua in which the Universe 
%could remain trapped during its thermal evolution all the way to $T \to 0$.

Next, one can aim at building a realistic model incorporating the above phase transition dynamics.
Using the results of this paper as guidelines, one can construct a 2HDM or a 3HDM with a softly broken symmetry group,
complete it with a suitable Yukawa sector, and use the finite-temperature loop-corrected effective potential
to check whether the regimes outlined here persist in the actual thermal history of the model.
Multi-step phase transitions, competing minima and saddle points, dynamics of charge-breaking bubble walls, 
charge-breaking, charge-restoring, or CB-to-CB phase transitions 
are the features to look for. The recent study \cite{Aoki:2023lbz} of the charge breaking vacuum 
at intermediate temperatures with all sorts of accompanying phase transitions proves than the one-loop effective potential
picture can be even richer than the tree-level one.

Finally, one should explore in depth whether the tantalizing possibility of charge-breaking bubble walls 
and charge-breaking intermediate phases has significant impact on baryogenesis, on gravitational wave spectrum,
and on the origin of primordial magnetic fields \cite{Durrer:2013pga,Subramanian:2015lua,Olea-Romacho:2023rhh}.
It may well happen that the intricate multi-Higgs dynamics discussed here opens novel opportunities 
for the cosmological history of the early Universe.

\section*{Acknowledgments}

This work was supported by Guangdong Natural Science Foundation (project No.~2024A1515012789).

\appendix

\section{The bilinear formalism in 3HDM}\label{appendix:bilinear}

The Higgs potential of the 3HDM depends on the three Higgs doublets $\phi_i$, $i=1,2,3$
via gauge-invariant bilinear combinations $\phi^\dagger_i \phi_j$, which define the gauge orbits
in the space of Higgs fields.
It allows one to recast the analysis of the potential in this bilinear space. 
A particularly useful version of the bilinear formalism was developed in mid-2000s first for the 2HDM, later for the NHDM, 
by three independent groups
\cite{Ivanov:2005hg,Nishi:2006tg,Maniatis:2006fs,Nishi:2007nh,Ivanov:2010ww,Ivanov:2010wz,Maniatis:2014oza}.
For the sake of completeness, we provide here details using the notation of \cite{Ivanov:2010ww}.

We begin by grouping the nine bilinears $\phi^\dagger_i \phi_j$ into scomponents of
a $1+8$-dimensional real vector
\begin{equation}
	r_0 = \frac{1}{\sqrt{3}}\sum_i \phi_i^\dagger \phi_i\,,\quad r_a = \sum_{i,j} \phi_i^\dagger \lambda^a_{ij}\phi_j\,,
	\quad a = 1, \dots, 8\,,
	\label{rmu}
\end{equation}
where $\lambda^a$ are the generators of $SU(3)$.
The explicit expressions are
\begin{eqnarray}
	&& r_0 = \frac{(\phi_1^\dagger\phi_1) + (\phi_2^\dagger\phi_2) + (\phi_3^\dagger\phi_3)}{\sqrt{3}}\,,\ 
	r_3 = \frac{(\phi_1^\dagger\phi_1) - (\phi_2^\dagger\phi_2)}{2}\,,\ 
	r_8 = \frac{(\phi_1^\dagger\phi_1) + (\phi_2^\dagger\phi_2) - 2(\phi_3^\dagger\phi_3)}{2\sqrt{3}} \quad
	\nonumber\\
	&&r_1 = \Re(\phi_1^\dagger\phi_2)\,,\quad 
	r_4 = \Re(\phi_3^\dagger\phi_1)\,,\quad 
	r_6 = \Re(\phi_2^\dagger\phi_3)\,,\nonumber\\[2mm] 
	&&r_2 = \Im(\phi_1^\dagger\phi_2)\,,\quad
	r_5 = \Im(\phi_3^\dagger\phi_1)\,,\quad
	r_7 = \Im(\phi_2^\dagger\phi_3)\,. \label{ri3HDM}
\end{eqnarray}
The 3HDM orbit space in the $1+8$-dimensional real space of bilinears $(r_0,r_a)$ is defined by
\begin{equation}
	r_0 \ge 0\,,\quad \vec r^{\,2} \le r_0^2\,,\quad \sqrt{3}d_{abc} r_a r_b r_c = \frac{3 \vec r^{\,2} - r_0^2}{2}r_0\,,
	\label{3HDMconditions}
\end{equation}
where $d_{abc}$ is the fully symmetric $SU(3)$ invariant tensor.
It is clear from the last equality, that, for a finite $r_0$, the modulus of the vector $\vec r$
cannot be arbitrarily small.
Let us denote $x = \vec r^{\,2}/r_0^2$.
Then the values of $x$ are restricted by $1/4 \le x \le 1$ \cite{Ivanov:2010ww}.
The neutral vevs always lie on the surface of the outer cone $\vec r^{\,2} = r_0^2$
and correspond to $x=1$. The charge-breaking extrema lie strictly inside the cone and correspond to $x < 1$. 
Thus, the departure of the quantity $x$ from 1 describes the ``charge-breaking depth'' of the point.

\section{Parameter dependence of extrema in the $\Sigma(36)$ 3HDM}\label{appendix:dependence}

\subsubsection{Numerical procedure}

In our numerical study, we find the extrema of the potential using the following procedure,
which we implemented in python and in {\tt Mathematica} to be able to cross check the outcomes.
\begin{itemize}
	\item We use linear parametrization of the three Higgs doublets, in which every complex scalar field
	is described as two real component. Overall, we get 12-dimensional space of real components $x_i$, $i = 1, \dots, 12$.
	\item
	We explicitly compute the gradient of the potential with respect to these components and 
	numerically solve the system of equations $\vec \nabla V = 0$ starting from a random seed point
	in the 12D space generated within a predefined box.
	\item
	When the solver finds a solution in the $x_i$ space, we use the bilinear formalism and compute
	the corresponding $1+8$ dimensional vector $(r_0, r_a)$, see Appendix~\ref{appendix:bilinear}.
	We store this solution in the form of $(r_0, r_a)$, not $x_i$.
	Whenever needed, we convert it from $(r_0, r_a)$ back to the three doublets, selecting a convenient 
	representative point in the gauge orbit. 
	\item
	A single run of the solver finds only one extremum.
	Staying with the same potential, we select another random seed point and run the solver again.
	When it converges, we check whether it represents a new solution by comparing its vector $(r_0, r_a)$
	with all other vectors already found and stored. The comparison is done within a certain tolerance level
	to account for numerical noise in floating point computations.
	\item
	At each new extremum, we compute the Hessian, find its eigenvalues, and store the results.
	The Hessian for a neutral minimum must possess three zero eigenvalues (the would-be Goldstone bosons)
	and nine positive eigenvalues which are the masses squared of the charged and neutral scalars. 
	A neutral saddle point contains positive and negative eigenvalues, in addition to the three zeros.
	For a charge breaking extremum, the multiplicity of the zero eigenvalue must increase to four.
	\item
	We repeat this procedure $N$ times, with a sufficiently large $N$ so that new extrema no longer appear 
	for any further seed point. 
	We found empirically that, for a typical set of the coefficients of the potential, we need several thousand
	seed points to detect all of its the extrema. In our calculations we used $N$ up to $10,\!000$,
	which took a few minutes of running time on a standard laptop.
	This large number of seed points looks unavoidable; limiting the search to one thousand seed points
	often misses a few extrema.
	\item
	As we change the coefficients of the potential, or add soft breaking terms, the extrema can shift,
	merge, or split. The depth of the potential at each point also changes. We found it necessary to track
	the evolution of all the extrema in order to have the good understanding of the global picture.
\end{itemize}

\subsubsection{Rigid vs. floating extrema}

\begin{figure}[!htb]
	\centering
	\includegraphics[width=0.48\textwidth]{sym-1-1-labels.pdf}
	%	\qquad 
	\includegraphics[width=0.48\textwidth]{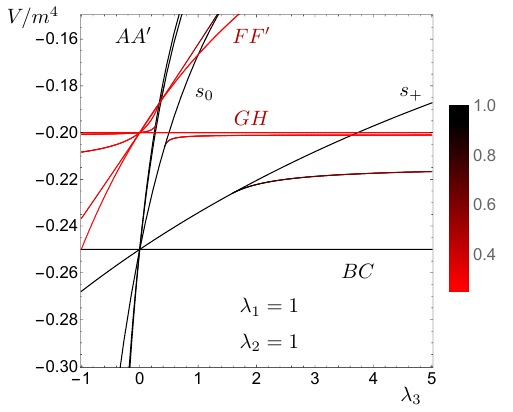}
	\caption{The reduced depths of all the extrema of the $\Sigma(36)$ 3HDM as a function of $\lambda_3$, computed for $\lambda_1 = 1$, $\lambda_2 = 1$. Left: the full picture. Right: a zoom on a narrow region of $V/m^4$.
		Color encodes the value of $x$, with the neutral extrema corresponding to $x=1$ (black)
		and the maximally charge-breaking extrema corresponding to $x=1/4$ (red).
		The labels indicate some of the rigid extrema, see the main text.}
	\label{fig-scan-1-again}
\end{figure}

In addition to the main text, we provide here more details on the non-trivial behavior of the extrema in the $\Sigma(36)$-symmetric 3HDM. 
To this end, we reproduce in Fig.~\ref{fig-scan-1-again} the left plot of Fig.~\ref{fig-scan-1} 
together with a zoom on the narrow range of $V/m^4$ where mergers and splittings of various extrema take place.

The first observation is that the vev alignments for some extrema are rigid (their vevs are calculable and are stable upon smooth
variation of $\lambda_3$), while the others are ``floating''.
Among the rigid alignments, we have the global minimum, which is attained at points $A$ and $A'$ for $-1 < \lambda_3 < 0$
and at points $B$ and $C$ for $\lambda_3 > 0$.
It turns out that all the special points mentioned in the previous section, both neutral ($A, A', B, C$) 
and charge-breaking ones ($F, F', G, H$), even when they are not minima, always represent rigid extrema of this potential.
Their vev alignments are not sensitive to the exact values of the coefficients, and the depths of the potential
$V/m^4$ at these points follow simple laws:
\begin{equation}
	BC: \  -\frac{1}{4\lambda_1}\,, \qquad
	AA': \  -\frac{1}{4(\lambda_1+\lambda_3)}\,, \qquad
	GH:\  -\frac{1}{4\lambda_1+\lambda_2}\,, \qquad
	FF': \ -\frac{1}{4\lambda_1+\lambda_2 + \lambda_3}\,.\label{rigid-depths-1}
\end{equation}
In addition, we found
other neutral rigid points which were not identified through the geometric procedure 
of the previous section because they lie inside the orbit space and never correspond to a global minimum.
The vev alignments and the values of $y$ of these extrema are
\begin{equation}
	s_0: \ (0,\, 1,\, -1)\,, \ y = \frac{1}{2},\quad 
	s_+: \ (1+\sqrt{3},\, 1,\, 1)\,, \ y = \frac{1}{2(\sqrt{3}+1)^2},\quad 
	s_-: \ (1-\sqrt{3},\, 1,\, 1)\,, \ y = \frac{1}{2(\sqrt{3}-1)^2}\,.\label{s+s0s-}
\end{equation}
Other points of the same type can be obtained by applying $\Sigma(36)$ 
transformations.
Notice that the extremum $s_-$ has the value $y \approx 0.933$, which is remarkably close to 1.
This explains why this curve follows so closely the vev alignment $AA'$, which corresponds to $y=1$.

In addition to the rigid points, we also observe in Fig.~\ref{fig-scan-1} other extrema, 
whose alignments and positions change smoothly as we vary $\lambda_3$.
All of these points, which we call the ``floating extrema'', lie in the charge-breaking part of the orbit space. 
Their ``charge breaking depth'', quantified by the parameter $1/4 \le x \le 1$ 
and shown in Fig.~\ref{fig-scan-1} with the shades of red color, varies smoothly as a function of $\lambda_3$.

Since we keep track, for each extremum, of the eigenvalues of the Hessian,
we observe that, for positive $\lambda_2$, only the previously found points $A$, $A'$, $B$, $C$
can be the minima. There are no other local minima in this potential; all other extrema are saddle points,
although with different signatures of the Hessian matrices. This allows us to conclude
that within the 3HDM with the exact $\Sigma(36)$ symmetry, no first order phase transitions are possible,
at least at the tree level.

The only phase transition which can take place for positive $\lambda_2$ is the transition 
from $A,A'$ to $B,C$ at the point $\lambda_3 = 0$, the value at which the $\Sigma(36)$ symmetry is 
accidentally restored to the full $SU(3)$. Although there is a large finite jump in vevs as we move through
this point, this is not a first order phase transition since the coexisting minima are not separated 
by a finite barrier.

\subsubsection{Bifurcation points and the number of extrema}

Unlike rigid extrema, which exist for all values of the parameters,
the floating extrema are present only within certain ranges of $\lambda_i$.
As we scan over $\lambda_3$, we observe them either merge with or emerge from a rigid neutral extremum.
A closer inspection shows that these bifurcations happen only along the curves corresponding 
to the rigid saddle points $s_0$, $s_+$, $s_-$, not the global minima.
We do not have the analytic expressions for the floating point location, 
but we do have them for the rigid points.
In order to find the exact instants of splitting or merging, 
we use the condition that, at the bifurcation point of the curves of $s_0$, $s_+$, $s_-$, 
one of the charged Higgs masses vanishes. 
The charged Higgs mass matrix can be computed analytically, 
thus we set its determinant to zero and obtain an equation for $\lambda_3$. 
Solutions of these equations give the values of $\lambda_3$ at which bifurcations take place.
We solved them with {\tt Mathematica} and found the following points: 
\begin{equation}
	\lambda_3^{(a)} = - \frac{2\lambda_2}{\sqrt{3}} \frac{\sqrt{3}+1}{\sqrt{3}-1}\,,\qquad
	\lambda_3^{(b)} = \frac{2\lambda_2}{\sqrt{3}} \frac{\sqrt{3}-1}{\sqrt{3}+1}\,,\qquad
	\lambda_3^{(c)} = \frac{2\lambda_2}{\sqrt{3}} \frac{1}{\sqrt{3}+1}\,,\qquad
	\lambda_3^{(d)} = \frac{2\lambda_2}{\sqrt{3}} \frac{1}{\sqrt{3}-1}\,,\label{mergers}
\end{equation}
ordered as $\lambda_3^{(a)} < \lambda_3^{(b)} < \lambda_3^{(c)} < \lambda_3^{(d)}$.
Note that the negative value $\lambda_3^{(a)}$ is available only when it satisfies the BFB conditions.

As we scan the entire $\lambda_3$ interval starting from the negative values,
we observe the following sequences of splittings and mergers:
\begin{itemize}
	\item The point $\lambda_3^{(a)}$ is not available for our choice $\lambda_2 = 1$, as it would violate the BFB conditions.
	\item At $\lambda_3 = \lambda_3^{(b)} \approx 0.31$, a floating extremum merges with $s_-$.
	\item At $\lambda_3 = \lambda_3^{(c)} \approx 0.42$, another floating extremum merges with $s_-$
	and, at the same time, a new floating extremum, located at a different depth, splits from $s_0$.
	\item At $\lambda_3 = \lambda_3^{(d)} \approx 1.58$, a floating extremum merges with $s_0$
	and, simultaneously, a new floating extremum, lying much deeper, splits from $s_-$.
\end{itemize}
Let us also count the total number of minima and of all extrema.
It is worth reminding the reader that, within the general 2HDM, the potential can admit at most two neutral minima 
plus four neutral and one charge-breaking saddle points, not counting the extremum at the origin.
Thus, the total number of non-trivial extrema in any 2HDM is at most seven.

Within the $\Sigma(36)$ 3HDM, we have a much richer structure of the potential, 
displayed by Fig.~\ref{fig-scan-1-again},
which unavoidably leads to a large number of coexisting extrema.
Notice that some of these extrema are linked by the transformations from $\Sigma(36)$
and, therefore, correspond to the same line in the above plots. 

The total number of minima always stays at six. These are the global minima identified earlier, 
which, for the exactly $\Sigma(36)$ symmetric case, are never accompanied with local minima.
By inspecting each rigid extremum and counting the broken and conserved symmetries, we find that 
the neutral points $A$, $A'$, $B$, $C$ give, together, 12 separate extrema,
six of which are always the global minima.
The same counting holds for the rigid charge-breaking points $F$, $F'$, $G$, $H$, which give 12 CB extrema.
The additional neutral rigid points, $s_+$, $s_-$, $s_0$, give nine additional extrema each, 27 in total.
Finally, each floating charge-breaking point is found to provide additional nine extrema.
There are two or three such floating points, depending on $\lambda_3$.
Thus, the total number of extrema $N_{\tiny \rm extrema}$ as a function of $\lambda_3$ is
\begin{equation}
	N_{\tiny \rm extrema} = 78\quad \mbox{for $\lambda_3 <  \lambda_3^{(b)} $}\,,
	\qquad
	N_{\tiny \rm extrema} = 69\quad \mbox{for $\lambda_3 >  \lambda_3^{(b)} $}\,.
\end{equation}
Note that the total number of extrema does not change when $\lambda_3$ passes through the other bifurcation points, 
$\lambda_3^{(c)}$ and $\lambda_3^{(d)}$: a merger of several extrema is accompanied by simultaneous
appearance of other extrema, although at a different depth.

We would like to repeat that, in order to spot all these extrema 
for a single value of $\lambda_3$, we had to run our code up to 10,000 times with random seed points.
Stopping the search at 1000 trials occasionally missed some of the extrema.
This technical challenge can further aggravate in models with even larger number of scalar fields.


\begin{thebibliography}{99}

%\cite{Branco:2011iw}
\bibitem{Branco:2011iw}
G.~C.~Branco, P.~M.~Ferreira, L.~Lavoura, M.~N.~Rebelo, M.~Sher and J.~P.~Silva,
%``Theory and phenomenology of two-Higgs-doublet models,''
Phys. Rept. \textbf{516}, 1-102 (2012)
doi:10.1016/j.physrep.2012.02.002
[arXiv:1106.0034 [hep-ph]].
%2723 citations counted in INSPIRE as of 10 Jan 2024

%\cite{Kanemura:2014bqa}
\bibitem{Kanemura:2014bqa}
S.~Kanemura, K.~Tsumura, K.~Yagyu and H.~Yokoya,
%``Fingerprinting nonminimal Higgs sectors,''
Phys. Rev. D \textbf{90}, 075001 (2014)
doi:10.1103/PhysRevD.90.075001
[arXiv:1406.3294 [hep-ph]].
%120 citations counted in INSPIRE as of 09 Jan 2024

%\cite{Ivanov:2017dad}
\bibitem{Ivanov:2017dad}
I.~P.~Ivanov,
%``Building and testing models with extended Higgs sectors,''
Prog. Part. Nucl. Phys. \textbf{95}, 160-208 (2017)
doi:10.1016/j.ppnp.2017.03.001
[arXiv:1702.03776 [hep-ph]].
%158 citations counted in INSPIRE as of 09 Jan 2024

%\cite{Arcadi:2019lka}
\bibitem{Arcadi:2019lka}
G.~Arcadi, A.~Djouadi and M.~Raidal,
%``Dark Matter through the Higgs portal,''
Phys. Rept. \textbf{842}, 1-180 (2020)
doi:10.1016/j.physrep.2019.11.003
[arXiv:1903.03616 [hep-ph]].
%229 citations counted in INSPIRE as of 09 Jan 2024

%\cite{Hindmarsh:2020hop}
\bibitem{Hindmarsh:2020hop}
M.~B.~Hindmarsh, M.~L\"uben, J.~Lumma and M.~Pauly,
%``Phase transitions in the early universe,''
SciPost Phys. Lect. Notes \textbf{24}, 1 (2021)
doi:10.21468/SciPostPhysLectNotes.24
[arXiv:2008.09136 [astro-ph.CO]].
%160 citations counted in INSPIRE as of 10 Jan 2024

%\cite{Athron:2023xlk}
\bibitem{Athron:2023xlk}
P.~Athron, C.~Bal\'azs, A.~Fowlie, L.~Morris and L.~Wu,
%``Cosmological phase transitions: From perturbative particle physics to gravitational waves,''
Prog. Part. Nucl. Phys. \textbf{135}, 104094 (2024)
doi:10.1016/j.ppnp.2023.104094
[arXiv:2305.02357 [hep-ph]].

%\cite{Ginzburg:2009dp}
\bibitem{Ginzburg:2009dp}
I.~F.~Ginzburg, I.~P.~Ivanov and K.~A.~Kanishev,
%``The Evolution of vacuum states and phase transitions in 2HDM during cooling of Universe,''
Phys. Rev. D \textbf{81}, 085031 (2010)
doi:10.1103/PhysRevD.81.085031
[arXiv:0911.2383 [hep-ph]].
%39 citations counted in INSPIRE as of 09 Jan 2024

%\cite{Ivanov:2008er}
\bibitem{Ivanov:2008er}
I.~P.~Ivanov,
%``Thermal evolution of the ground state of the most general 2HDM,''
Acta Phys. Polon. B \textbf{40}, 2789-2807 (2009)
[arXiv:0812.4984 [hep-ph]].
%32 citations counted in INSPIRE as of 09 Jan 2024

%\cite{Ginzburg:2010wa}
\bibitem{Ginzburg:2010wa}
I.~F.~Ginzburg, K.~A.~Kanishev, M.~Krawczyk and D.~Sokolowska,
%``Evolution of Universe to the present inert phase,''
Phys. Rev. D \textbf{82}, 123533 (2010)
doi:10.1103/PhysRevD.82.123533
[arXiv:1009.4593 [hep-ph]].
%138 citations counted in INSPIRE as of 09 Jan 2024

%\cite{Vaskonen:2016yiu}
\bibitem{Vaskonen:2016yiu}
V.~Vaskonen,
%``Electroweak baryogenesis and gravitational waves from a real scalar singlet,''
Phys. Rev. D \textbf{95}, no.12, 123515 (2017)
doi:10.1103/PhysRevD.95.123515
[arXiv:1611.02073 [hep-ph]].
%173 citations counted in INSPIRE as of 10 Jan 2024

%\cite{Chala:2018opy}
\bibitem{Chala:2018opy}
M.~Chala, M.~Ramos and M.~Spannowsky,
%``Gravitational wave and collider probes of a triplet Higgs sector with a low cutoff,''
Eur. Phys. J. C \textbf{79}, no.2, 156 (2019)
doi:10.1140/epjc/s10052-019-6655-1
[arXiv:1812.01901 [hep-ph]].
%56 citations counted in INSPIRE as of 10 Jan 2024

%\cite{Morais:2019fnm}
\bibitem{Morais:2019fnm}
A.~P.~Morais and R.~Pasechnik,
%``Probing multi-step electroweak phase transition with multi-peaked primordial gravitational waves spectra,''
JCAP \textbf{04}, 036 (2020)
doi:10.1088/1475-7516/2020/04/036
[arXiv:1910.00717 [hep-ph]].
%22 citations counted in INSPIRE as of 10 Jan 2024

%\cite{Aoki:2021oez}
\bibitem{Aoki:2021oez}
M.~Aoki, T.~Komatsu and H.~Shibuya,
%``Possibility of a multi-step electroweak phase transition in the two-Higgs doublet models,''
PTEP \textbf{2022}, no.6, 063B05 (2022)
doi:10.1093/ptep/ptac068
[arXiv:2106.03439 [hep-ph]].
%22 citations counted in INSPIRE as of 10 Jan 2024

%\cite{Benincasa:2022elt}
\bibitem{Benincasa:2022elt}
N.~Benincasa, L.~Delle Rose, K.~Kannike and L.~Marzola,
%``Multi-step phase transitions and gravitational waves in the inert doublet model,''
JCAP \textbf{12}, 025 (2022)
doi:10.1088/1475-7516/2022/12/025
[arXiv:2205.06669 [hep-ph]].
%17 citations counted in INSPIRE as of 10 Jan 2024

%\cite{Cao:2022ocg}
\bibitem{Cao:2022ocg}
Q.~H.~Cao, K.~Hashino, X.~X.~Li and J.~H.~Yue,
%``Multi-step phase transition and gravitational wave from general $\mathbb{Z}_2$ scalar extensions,''
[arXiv:2212.07756 [hep-ph]].
%6 citations counted in INSPIRE as of 10 Jan 2024

%\cite{Aoki:2023xnn}
\bibitem{Aoki:2023xnn}
M.~Aoki and H.~Shibuya,
%``Electroweak baryogenesis between broken phases in multi-step phase transition,''
Phys. Lett. B \textbf{843}, 138041 (2023)
doi:10.1016/j.physletb.2023.138041
[arXiv:2302.11551 [hep-ph]].
%2 citations counted in INSPIRE as of 09 Jan 2024

%\cite{Aoki:2023lbz}
\bibitem{Aoki:2023lbz}
M.~Aoki, L.~Biermann, C.~Borschensky, I.~P.~Ivanov, M.~M\"uhlleitner and H.~Shibuya,
%``Intermediate charge-breaking phases and symmetry non-restoration in the 2-Higgs-Doublet Model,''
JHEP \textbf{02}, 232 (2024)
doi:10.1007/JHEP02(2024)232
[arXiv:2308.04141 [hep-ph]].

\bibitem{Basler:2024aaf}
P.~Basler, L.~Biermann, M.~M\"uhlleitner, J.~M\"uller, R.~Santos and J.~Viana,
%``BSMPT v3 A Tool for Phase Transitions and Primordial Gravitational Waves in Extended Higgs Sectors,''
[arXiv:2404.19037 [hep-ph]].
%0 citations counted in INSPIRE as of 31 May 2024

%\cite{Weinberg:1974hy}
\bibitem{Weinberg:1974hy}
S.~Weinberg,
%``Gauge and Global Symmetries at High Temperature,''
Phys. Rev. D \textbf{9}, 3357-3378 (1974)
doi:10.1103/PhysRevD.9.3357
%1564 citations counted in INSPIRE as of 10 Jan 2024

%\cite{Biekotter:2021ysx}
\bibitem{Biekotter:2021ysx}
T.~Biek\"otter, S.~Heinemeyer, J.~M.~No, M.~O.~Olea and G.~Weiglein,
%``Fate of electroweak symmetry in the early Universe: Non-restoration and trapped vacua in the N2HDM,''
JCAP \textbf{06}, 018 (2021)
doi:10.1088/1475-7516/2021/06/018
[arXiv:2103.12707 [hep-ph]].
%36 citations counted in INSPIRE as of 10 Jan 2024

\bibitem{Biekotter:2022kgf}
T.~Biek\"otter, S.~Heinemeyer, J.~M.~No, M.~O.~Olea-Romacho and G.~Weiglein,
%``The trap in the early Universe: impact on the interplay between gravitational waves and LHC physics in the 2HDM,''
JCAP \textbf{03}, 031 (2023)
doi:10.1088/1475-7516/2023/03/031
[arXiv:2208.14466 [hep-ph]].

%\cite{Barroso:2006pa}
\bibitem{Barroso:2006pa}
A.~Barroso, P.~M.~Ferreira, R.~Santos and J.~P.~Silva,
%``Stability of the normal vacuum in multi-Higgs-doublet models,''
Phys. Rev. D \textbf{74}, 085016 (2006)
doi:10.1103/PhysRevD.74.085016
[arXiv:hep-ph/0608282 [hep-ph]].
%63 citations counted in INSPIRE as of 09 Jan 2024

%\cite{Ginzburg:2007jn}
\bibitem{Ginzburg:2007jn}
I.~F.~Ginzburg and K.~A.~Kanishev,
%``Different vacua in 2HDM,''
Phys. Rev. D \textbf{76}, 095013 (2007)
doi:10.1103/PhysRevD.76.095013
[arXiv:0704.3664 [hep-ph]].
%38 citations counted in INSPIRE as of 09 Jan 2024

%\cite{Basler:2020nrq}
\bibitem{Basler:2020nrq}
P.~Basler, M.~M\"uhlleitner and J.~M\"uller,
%``BSMPT v2 a tool for the electroweak phase transition and the baryon asymmetry of the universe in extended Higgs Sectors,''
Comput. Phys. Commun. \textbf{269}, 108124 (2021)
doi:10.1016/j.cpc.2021.108124
[arXiv:2007.01725 [hep-ph]].
%39 citations counted in INSPIRE as of 09 Jan 2024

%\cite{Zeldovich:1974uw}
\bibitem{Zeldovich:1974uw}
Y.~B.~Zeldovich, I.~Y.~Kobzarev and L.~B.~Okun,
%``Cosmological Consequences of the Spontaneous Breakdown of Discrete Symmetry,''
Zh. Eksp. Teor. Fiz. \textbf{67}, 3-11 (1974)
SLAC-TRANS-0165.
%960 citations counted in INSPIRE as of 10 Jan 2024

%\cite{Kibble:1976sj}
\bibitem{Kibble:1976sj}
T.~W.~B.~Kibble,
%``Topology of Cosmic Domains and Strings,''
J. Phys. A \textbf{9}, 1387-1398 (1976)
doi:10.1088/0305-4470/9/8/029
%3156 citations counted in INSPIRE as of 10 Jan 2024

%\cite{Blasi:2022woz}
\bibitem{Blasi:2022woz}
S.~Blasi and A.~Mariotti,
%``Domain Walls Seeding the Electroweak Phase Transition,''
Phys. Rev. Lett. \textbf{129}, no.26, 261303 (2022)
doi:10.1103/PhysRevLett.129.261303
[arXiv:2203.16450 [hep-ph]].
%22 citations counted in INSPIRE as of 10 Jan 2024

%\cite{Agrawal:2023cgp}
\bibitem{Agrawal:2023cgp}
P.~Agrawal, S.~Blasi, A.~Mariotti and M.~Nee,
%``Electroweak Phase Transition with a Double Well Done Doubly Well,''
[arXiv:2312.06749 [hep-ph]].
%0 citations counted in INSPIRE as of 09 Jan 2024

%\cite{Battye:2020sxy}
\bibitem{Battye:2020sxy}
R.~A.~Battye, A.~Pilaftsis and D.~G.~Viatic,
%``Simulations of Domain Walls in Two Higgs Doublet Models,''
JHEP \textbf{01}, 105 (2021)
doi:10.1007/JHEP01(2021)105
[arXiv:2006.13273 [hep-ph]].
%11 citations counted in INSPIRE as of 10 Jan 2024

%\cite{Battye:2020jeu}
\bibitem{Battye:2020jeu}
R.~A.~Battye, A.~Pilaftsis and D.~G.~Viatic,
%``Domain wall constraints on two-Higgs-doublet models with $Z_2$ symmetry,''
Phys. Rev. D \textbf{102}, no.12, 123536 (2020)
doi:10.1103/PhysRevD.102.123536
[arXiv:2010.09840 [hep-ph]].
%14 citations counted in INSPIRE as of 10 Jan 2024

%\cite{Sassi:2023cqp}
\bibitem{Sassi:2023cqp}
M.~Y.~Sassi and G.~Moortgat-Pick,
%``Domain walls in the Two-Higgs-Doublet Model and their charge and CP-violating interactions with Standard Model fermions,''
JHEP \textbf{04}, 101 (2024)
doi:10.1007/JHEP04(2024)101
[arXiv:2309.12398 [hep-ph]].

%\cite{Durrer:2013pga}
\bibitem{Durrer:2013pga}
R.~Durrer and A.~Neronov,
%``Cosmological Magnetic Fields: Their Generation, Evolution and Observation,''
Astron. Astrophys. Rev. \textbf{21}, 62 (2013)
doi:10.1007/s00159-013-0062-7
[arXiv:1303.7121 [astro-ph.CO]].
%623 citations counted in INSPIRE as of 09 Jan 2024

%\cite{Subramanian:2015lua}
\bibitem{Subramanian:2015lua}
K.~Subramanian,
%``The origin, evolution and signatures of primordial magnetic fields,''
Rept. Prog. Phys. \textbf{79}, no.7, 076901 (2016)
doi:10.1088/0034-4885/79/7/076901
[arXiv:1504.02311 [astro-ph.CO]].
%364 citations counted in INSPIRE as of 09 Jan 2024

%\cite{Olea-Romacho:2023rhh}
\bibitem{Olea-Romacho:2023rhh}
M.~O.~Olea-Romacho,
%``Primordial magnetogenesis in the two-Higgs-doublet model,''
Phys. Rev. D \textbf{109}, no.1, 015023 (2024)
doi:10.1103/PhysRevD.109.015023
[arXiv:2310.19948 [hep-ph]].


%\cite{Weinberg:1976hu}
\bibitem{Weinberg:1976hu}
S.~Weinberg,
%``Gauge Theory of CP Violation,''
Phys. Rev. Lett. \textbf{37}, 657 (1976)
doi:10.1103/PhysRevLett.37.657

%\cite{Ivanov:2010wz}
\bibitem{Ivanov:2010wz}
I.~P.~Ivanov,
%``Properties of the general NHDM. II. Higgs potential and its symmetries,''
JHEP \textbf{07}, 020 (2010)
doi:10.1007/JHEP07(2010)020
[arXiv:1004.1802 [hep-th]].
%37 citations counted in INSPIRE as of 09 Jan 2024

%\cite{Ivanov:2012fp}
\bibitem{Ivanov:2012fp}
I.~P.~Ivanov and E.~Vdovin,
%``Classification of finite reparametrization symmetry groups in the three-Higgs-doublet model,''
Eur. Phys. J. C \textbf{73}, no.2, 2309 (2013)
doi:10.1140/epjc/s10052-013-2309-x
[arXiv:1210.6553 [hep-ph]].
%94 citations counted in INSPIRE as of 09 Jan 2024

%\cite{Ivanov:2012ry}
\bibitem{Ivanov:2012ry}
I.~P.~Ivanov and E.~Vdovin,
%``Discrete symmetries in the three-Higgs-doublet model,''
Phys. Rev. D \textbf{86}, 095030 (2012)
doi:10.1103/PhysRevD.86.095030
[arXiv:1206.7108 [hep-ph]].
%54 citations counted in INSPIRE as of 09 Jan 2024

%\cite{Ivanov:2014doa}
\bibitem{Ivanov:2014doa}
I.~P.~Ivanov and C.~C.~Nishi,
%``Symmetry breaking patterns in 3HDM,''
JHEP \textbf{01}, 021 (2015)
doi:10.1007/JHEP01(2015)021
[arXiv:1410.6139 [hep-ph]].
%78 citations counted in INSPIRE as of 09 Jan 2024

%\cite{Segre:1978ji}
\bibitem{Segre:1978ji}
G.~Segre and H.~A.~Weldon,
%``Mass Hierarchies and a Formula for the Cabibbo Angle in SU(2)$_L \times$ U(1),''
Phys. Lett. B \textbf{83}, 351-354 (1979)
doi:10.1016/0370-2693(79)91125-0
%42 citations counted in INSPIRE as of 09 Jan 2024

%\cite{Grimus:2010ak}
\bibitem{Grimus:2010ak}
W.~Grimus and P.~O.~Ludl,
%``Principal series of finite subgroups of SU(3),''
J. Phys. A \textbf{43}, 445209 (2010)
doi:10.1088/1751-8113/43/44/445209
[arXiv:1006.0098 [hep-ph]].
%51 citations counted in INSPIRE as of 09 Jan 2024

%\cite{Merle:2011vy}
\bibitem{Merle:2011vy}
A.~Merle and R.~Zwicky,
%``Explicit and spontaneous breaking of SU(3) into its finite subgroups,''
JHEP \textbf{02}, 128 (2012)
doi:10.1007/JHEP02(2012)128
[arXiv:1110.4891 [hep-ph]].
%63 citations counted in INSPIRE as of 09 Jan 2024

%\cite{Hagedorn:2013nra}
\bibitem{Hagedorn:2013nra}
C.~Hagedorn, A.~Meroni and L.~Vitale,
%``Mixing patterns from the groups $\Sigma(n\phi)$,''
J. Phys. A \textbf{47}, 055201 (2014)
doi:10.1088/1751-8113/47/5/055201
[arXiv:1307.5308 [hep-ph]].
%42 citations counted in INSPIRE as of 09 Jan 2024

%\cite{Rong:2016cpk}
\bibitem{Rong:2016cpk}
S.~j.~Rong,
%``Lepton mixing patterns from the group \ensuremath{\Sigma}(36\texttimes{}3) with a generalized CP transformation,''
Phys. Rev. D \textbf{95}, no.7, 076014 (2017)
doi:10.1103/PhysRevD.95.076014
[arXiv:1604.08482 [hep-ph]].
%21 citations counted in INSPIRE as of 09 Jan 2024

%\cite{deMedeirosVarzielas:2021zqs}
\bibitem{deMedeirosVarzielas:2021zqs}
I.~de Medeiros Varzielas, I.~P.~Ivanov and M.~Levy,
%``Exploring multi-Higgs models with softly broken large discrete symmetry groups,''
Eur. Phys. J. C \textbf{81}, no.10, 918 (2021)
doi:10.1140/epjc/s10052-021-09681-w
[arXiv:2107.08227 [hep-ph]].
%9 citations counted in INSPIRE as of 09 Jan 2024

%\cite{Ivanov:2011ae}
\bibitem{Ivanov:2011ae}
I.~P.~Ivanov, V.~Keus and E.~Vdovin,
%``Abelian symmetries in multi-Higgs-doublet models,''
J. Phys. A \textbf{45}, 215201 (2012)
doi:10.1088/1751-8113/45/21/215201
[arXiv:1112.1660 [math-ph]].
%81 citations counted in INSPIRE as of 09 Jan 2024

%\cite{Branco:1983tn}
\bibitem{Branco:1983tn}
G.~C.~Branco, J.~M.~Gerard and W.~Grimus,
%``GEOMETRICAL T VIOLATION,''
Phys. Lett. B \textbf{136}, 383-386 (1984)
doi:10.1016/0370-2693(84)92024-0
%157 citations counted in INSPIRE as of 09 Jan 2024

%\cite{Degee:2012sk}
\bibitem{Degee:2012sk}
A.~Degee, I.~P.~Ivanov and V.~Keus,
%``Geometric minimization of highly symmetric potentials,''
JHEP \textbf{02}, 125 (2013)
doi:10.1007/JHEP02(2013)125
[arXiv:1211.4989 [hep-ph]].
%53 citations counted in INSPIRE as of 09 Jan 2024

%\cite{deMedeirosVarzielas:2022fnn}
\bibitem{deMedeirosVarzielas:2022fnn}
I.~de Medeiros Varzielas and D.~Ivo,
%``Geometric minimization of softly broken potentials,''
Eur. Phys. J. Plus \textbf{138}, no.1, 81 (2023)
doi:10.1140/epjp/s13360-023-03670-6
[arXiv:2206.03577 [hep-ph]].
%1 citations counted in INSPIRE as of 09 Jan 2024

%\cite{Ivanov:2010ww}
\bibitem{Ivanov:2010ww}
I.~P.~Ivanov and C.~C.~Nishi,
%``Properties of the general NHDM. I. The Orbit space,''
Phys. Rev. D \textbf{82}, 015014 (2010)
doi:10.1103/PhysRevD.82.015014
[arXiv:1004.1799 [hep-th]].
%44 citations counted in INSPIRE as of 09 Jan 2024

%\cite{deMedeirosVarzielas:2011zw}
\bibitem{deMedeirosVarzielas:2011zw}
I.~de Medeiros Varzielas and D.~Emmanuel-Costa,
%``Geometrical CP Violation,''
Phys. Rev. D \textbf{84}, 117901 (2011)
doi:10.1103/PhysRevD.84.117901
[arXiv:1106.5477 [hep-ph]].
%77 citations counted in INSPIRE as of 09 Jan 2024

%\cite{Varzielas:2012nn}
\bibitem{Varzielas:2012nn}
I.~de Medeiros Varzielas, D.~Emmanuel-Costa and P.~Leser,
%``Geometrical CP Violation from Non-Renormalisable Scalar Potentials,''
Phys. Lett. B \textbf{716}, 193-196 (2012)
doi:10.1016/j.physletb.2012.08.008
[arXiv:1204.3633 [hep-ph]].
%72 citations counted in INSPIRE as of 09 Jan 2024

%\cite{Ivanov:2013nla}
\bibitem{Ivanov:2013nla}
I.~P.~Ivanov and L.~Lavoura,
%``Geometrical CP violation in the N-Higgs-doublet model,''
Eur. Phys. J. C \textbf{73}, no.4, 2416 (2013)
doi:10.1140/epjc/s10052-013-2416-8
[arXiv:1302.3656 [hep-ph]].
%29 citations counted in INSPIRE as of 09 Jan 2024

%\cite{Bhattacharyya:2012pi}
\bibitem{Bhattacharyya:2012pi}
G.~Bhattacharyya, I.~de Medeiros Varzielas and P.~Leser,
%``A common origin of fermion mixing and geometrical CP violation, and its test through Higgs physics at the LHC,''
Phys. Rev. Lett. \textbf{109}, 241603 (2012)
doi:10.1103/PhysRevLett.109.241603
[arXiv:1210.0545 [hep-ph]].
%89 citations counted in INSPIRE as of 09 Jan 2024

%\cite{Varzielas:2013sla}
\bibitem{Varzielas:2013sla}
I.~de Medeiros Varzielas and D.~Pidt,
%``Towards realistic models of quark masses with geometrical CP violation,''
J. Phys. G \textbf{41}, 025004 (2014)
doi:10.1088/0954-3899/41/2/025004
[arXiv:1307.0711 [hep-ph]].
%51 citations counted in INSPIRE as of 09 Jan 2024

%\cite{Varzielas:2013eta}
\bibitem{Varzielas:2013eta}
I.~de Medeiros Varzielas and D.~Pidt,
%``Geometrical CP violation with a complete fermion sector,''
JHEP \textbf{11}, 206 (2013)
doi:10.1007/JHEP11(2013)206
[arXiv:1307.6545 [hep-ph]].
%30 citations counted in INSPIRE as of 09 Jan 2024

%\cite{Fallbacher:2015rea}
\bibitem{Fallbacher:2015rea}
M.~Fallbacher and A.~Trautner,
%``Symmetries of symmetries and geometrical CP violation,''
Nucl. Phys. B \textbf{894}, 136-160 (2015)
doi:10.1016/j.nuclphysb.2015.03.003
[arXiv:1502.01829 [hep-ph]].
%29 citations counted in INSPIRE as of 09 Jan 2024

%\cite{deMedeirosVarzielas:2022kbj}
\bibitem{deMedeirosVarzielas:2022kbj}
I.~de Medeiros Varzielas and D.~Ivo,
%``Softly-broken $A_4$ or $S_4$ 3HDMs with stable states,''
Eur. Phys. J. C \textbf{82}, no.5, 415 (2022)
doi:10.1140/epjc/s10052-022-10331-y
[arXiv:2202.00681 [hep-ph]].
%7 citations counted in INSPIRE as of 09 Jan 2024

%\cite{Hagedorn:2023mrg}
\bibitem{Hagedorn:2023mrg}
C.~Hagedorn, M.~L.~L\'opez-Ib\'a\~nez, M.~J.~P\'erez, M.~H.~Rahat and O.~Vives,
%``Flavon vacuum alignment beyond SUSY,''
[arXiv:2312.07430 [hep-ph]].
%1 citations counted in INSPIRE as of 09 Jan 2024

%\cite{Ivanov:2005hg}
\bibitem{Ivanov:2005hg}
I.~P.~Ivanov,
%``Two-Higgs-doublet model from the group-theoretic perspective,''
Phys. Lett. B \textbf{632}, 360-365 (2006)
doi:10.1016/j.physletb.2005.10.015
[arXiv:hep-ph/0507132 [hep-ph]].
%88 citations counted in INSPIRE as of 09 Jan 2024

%\cite{Nishi:2006tg}
\bibitem{Nishi:2006tg}
C.~C.~Nishi,
%``CP violation conditions in N-Higgs-doublet potentials,''
Phys. Rev. D \textbf{74}, 036003 (2006)
[erratum: Phys. Rev. D \textbf{76}, 119901 (2007)]
doi:10.1103/PhysRevD.76.119901
[arXiv:hep-ph/0605153 [hep-ph]].
%132 citations counted in INSPIRE as of 09 Jan 2024

%\cite{Maniatis:2007vn}
\bibitem{Maniatis:2007vn}
M.~Maniatis, A.~von Manteuffel and O.~Nachtmann,
%``CP violation in the general two-Higgs-doublet model: A Geometric view,''
Eur. Phys. J. C \textbf{57}, 719-738 (2008)
doi:10.1140/epjc/s10052-008-0712-5
[arXiv:0707.3344 [hep-ph]].
%100 citations counted in INSPIRE as of 09 Jan 2024

%\cite{Maniatis:2007de}
\bibitem{Maniatis:2007de}
M.~Maniatis, A.~von Manteuffel and O.~Nachtmann,
%``A New type of CP symmetry, family replication and fermion mass hierarchies,''
Eur. Phys. J. C \textbf{57}, 739-762 (2008)
doi:10.1140/epjc/s10052-008-0726-z
[arXiv:0711.3760 [hep-ph]].
%45 citations counted in INSPIRE as of 09 Jan 2024

%\cite{Ferreira:2010yh}
\bibitem{Ferreira:2010yh}
P.~M.~Ferreira, H.~E.~Haber, M.~Maniatis, O.~Nachtmann and J.~P.~Silva,
%``Geometric picture of generalized-CP and Higgs-family transformations in the two-Higgs-doublet model,''
Int. J. Mod. Phys. A \textbf{26}, 769-808 (2011)
doi:10.1142/S0217751X11051494
[arXiv:1010.0935 [hep-ph]].
%59 citations counted in INSPIRE as of 09 Jan 2024

%\cite{Rubakov:2002fi}
\bibitem{Rubakov:2002fi}
V.~A.~Rubakov,
%``Classical theory of gauge fields,''
Princeton University Press, 2002,
ISBN 978-0-691-05927-3, 978-0-691-05927-3
%45 citations counted in INSPIRE as of 09 Jan 2024

%\cite{Coleman:1977py}
\bibitem{Coleman:1977py}
S.~R.~Coleman,
%``The Fate of the False Vacuum. 1. Semiclassical Theory,''
Phys. Rev. D \textbf{15}, 2929-2936 (1977)
[erratum: Phys. Rev. D \textbf{16}, 1248 (1977)]
doi:10.1103/PhysRevD.16.1248
%2524 citations counted in INSPIRE as of 10 Jan 2024


\bibitem{Dolan:1973qd}
L.~Dolan and R.~Jackiw,
%``Symmetry Behavior at Finite Temperature,''
Phys. Rev. D \textbf{9}, 3320-3341 (1974)
doi:10.1103/PhysRevD.9.3320

%\cite{Cline:1996mga}
\bibitem{Cline:1996mga}
J.~M.~Cline and P.~A.~Lemieux,
%``Electroweak phase transition in two Higgs doublet models,''
Phys. Rev. D \textbf{55}, 3873-3881 (1997)
doi:10.1103/PhysRevD.55.3873
[arXiv:hep-ph/9609240 [hep-ph]].
%206 citations counted in INSPIRE as of 09 Jan 2024

%\cite{Maniatis:2006fs}
\bibitem{Maniatis:2006fs}
M.~Maniatis, A.~von Manteuffel, O.~Nachtmann and F.~Nagel,
%``Stability and symmetry breaking in the general two-Higgs-doublet model,''
Eur. Phys. J. C \textbf{48}, 805-823 (2006)
doi:10.1140/epjc/s10052-006-0016-6
[arXiv:hep-ph/0605184 [hep-ph]].
%222 citations counted in INSPIRE as of 09 Jan 2024

%\cite{Nishi:2007nh}
\bibitem{Nishi:2007nh}
C.~C.~Nishi,
%``The Structure of potentials with N Higgs doublets,''
Phys. Rev. D \textbf{76}, 055013 (2007)
doi:10.1103/PhysRevD.76.055013
[arXiv:0706.2685 [hep-ph]].
%60 citations counted in INSPIRE as of 09 Jan 2024

%\cite{Maniatis:2014oza}
\bibitem{Maniatis:2014oza}
M.~Maniatis and O.~Nachtmann,
%``Stability and symmetry breaking in the general three-Higgs-doublet model,''
JHEP \textbf{02}, 058 (2015)
[erratum: JHEP \textbf{10}, 149 (2015)]
doi:10.1007/JHEP10(2015)149
[arXiv:1408.6833 [hep-ph]].
%36 citations counted in INSPIRE as of 09 Jan 2024
\end{thebibliography}
\end{document}